\documentstyle[epsfig]{mn}

\newif\ifAMStwofonts

\ifoldfss
  \ifCUPmtlplainloaded \else
    \NewTextAlphabet{textbfit} {cmbxti10} {}
    \NewTextAlphabet{textbfss} {cmssbx10} {}
    \NewMathAlphabet{mathbfit} {cmbxti10} {} 
    \NewMathAlphabet{mathbfss} {cmssbx10} {} 
  \fi
  \ifAMStwofonts
    \ifCUPmtlplainloaded \else
      \NewSymbolFont{upmath} {eurm10}
      \NewSymbolFont{AMSa} {msam10}
      \NewMathSymbol{\upi}     {0}{upmath}{19}
      \NewMathSymbol{\umu}     {0}{upmath}{16}
      \NewMathSymbol{\upartial}{0}{upmath}{40}
      \NewMathSymbol{\leqslant}{3}{AMSa}{36}
      \NewMathSymbol{\geqslant}{3}{AMSa}{3E}

       \let\le=\leqslant
       
    \fi
  \fi
\fi 

\ifnfssone
  \newmathalphabet{\mathit}
  \addtoversion{normal}{\mathit}{cmr}{m}{it}
  \addtoversion{bold}{\mathit}{cmr}{bx}{it}
  \newmathalphabet{\mathbfit} 
  \addtoversion{normal}{\mathbfit}{cmr}{bx}{it}
  \addtoversion{bold}{\mathbfit}{cmr}{bx}{it}
  \newmathalphabet{\mathbfss} 
  \addtoversion{normal}{\mathbfss}{cmss}{bx}{n}
  \addtoversion{bold}{\mathbfss}{cmss}{bx}{n}
  \ifAMStwofonts
    \ifCUPmtlplainloaded \else
      \UseAMStwoboldmath
      \makeatletter
      \new@mathgroup\upmath@group
      \define@mathgroup\mv@normal\upmath@group{eur}{m}{n}
      \define@mathgroup\mv@bold\upmath@group{eur}{b}{n}
      \edef\UPM{\hexnumber\upmath@group}
      \new@mathgroup\amsa@group
      \define@mathgroup\mv@normal\amsa@group{msa}{m}{n}
      \define@mathgroup\mv@bold\amsa@group{msa}{m}{n}
      \edef\AMSa{\hexnumber\amsa@group}
      \makeatother
      \mathchardef\upi="0\UPM19
      \mathchardef\umu="0\UPM16
      \mathchardef\upartial="0\UPM40
      \mathchardef\leqslant="3\AMSa36
      \mathchardef\geqslant="3\AMSa3E

       \let\le=\leqslant

    \fi
  \fi
\fi 

\ifnfsstwo
  \DeclareMathAlphabet{\mathbfit}{OT1}{cmr}{bx}{it}
  \SetMathAlphabet\mathbfit{bold}{OT1}{cmr}{bx}{it}
  \DeclareMathAlphabet{\mathbfss}{OT1}{cmss}{bx}{n}
  \SetMathAlphabet\mathbfss{bold}{OT1}{cmss}{bx}{n}
  \ifAMStwofonts
    \ifCUPmtlplainloaded \else
      \DeclareSymbolFont{UPM}{U}{eur}{m}{n}
      \SetSymbolFont{UPM}{bold}{U}{eur}{b}{n}
      \DeclareSymbolFont{AMSa}{U}{msa}{m}{n}
      \DeclareMathSymbol{\upi}{0}{UPM}{"19}
      \DeclareMathSymbol{\umu}{0}{UPM}{"16}
      \DeclareMathSymbol{\upartial}{0}{UPM}{"40}
      \DeclareMathSymbol{\leqslant}{3}{AMSa}{"36}
      \DeclareMathSymbol{\geqslant}{3}{AMSa}{"3E}

       \let\le=\leqslant

    \fi
  \fi
\fi 

\ifCUPmtlplainloaded \else
  \ifAMStwofonts \else 
    \def\upi{\pi}
    \def\umu{\mu}
    \def\upartial{\partial}
  \fi
\fi

\def\lesssim{\mathrel{\hbox{\rlap{\hbox{\lower4pt\hbox{$\sim$}}}\hbox{$<$}}}}
\def\gtrsim{\mathrel{\hbox{\rlap{\hbox{\lower4pt\hbox{$\sim$}}}\hbox{$>$}}}}
\newcommand{\ltsima}{$\; \buildrel < \over \sim \;$}
\newcommand{\simlt}{\lower.5ex\hbox{\ltsima}}
\newcommand{\gtsima}{$\; \buildrel > \over \sim \;$}
\newcommand{\simgt}{\lower.5ex\hbox{\gtsima}}

\title{The dust content of QSO hosts at high redshift} 
  \author[F. Calura et al.]  {F. Calura$^{1}$\thanks{E-mail:
      fcalura@oabo.inaf.it}, R. Gilli$^{1}$, C. Vignali$^{1,2}$, F. Pozzi$^{2}$, 
A. Pipino$^{3}$, F. Matteucci$^{4}$\\
(1) INAF, Osservatorio Astronomico 
di Bologna, Via Ranzani 1, 40127 Bologna, Italy\\ 
(2) Dipartimento di Fisica e Astronomia, Universit\'a degli Studi di Bologna, Viale Berti Pichat 6/2, 40127 Bologna\\
(3) Institut fur Astronomie, ETH Zurich, Wolfgang-Pauli-Str. 27, 8093, Zurich, Switzerland\\
(4) Dipartimento di
Fisica - Sezione di Astronomia, Universit\`a di Trieste, Via
G. B. Tiepolo 11, 34131 Trieste, Italy\\}

\pagerange{\pageref{firstpage}--\pageref{lastpage}}
\pubyear{---}

\begin{document}

\maketitle

\label{firstpage}
\begin{abstract}
Infrared observations of high-$z$ quasar (QSO) hosts indicate the presence of large masses 
of dust in the early universe. 
When combined with other observables, such as neutral gas masses and star formation rates, 
the dust content of $z\sim 6$ QSO hosts may help constraining their star formation history. 
We have collected a database of 58 sources from the literature discovered by various surveys and observed in the FIR. 
We have interpreted the available data by means of chemical evolution models 
for forming proto-spheroids, investigating the role of the major  
parameters regulating star formation and dust production. 
For a few systems, given the derived small dynamical masses, 
the observed dust content can be explained only assuming a top-heavy initial mass function, 
an enhanced star formation efficiency and an increased rate of dust accretion. 
However, the possibility that, for some systems, the dynamical mass has been underestimated cannot be excluded. 
If this were the case, the dust mass can be accounted for by standard model assumptions. 
We provide predictions regarding the abundance of the descendants of QSO hosts; 
albeit rare, such systems 
should be present and detectable by future deep surveys such as Euclid already at $z>4$. 
\end{abstract} 

\begin{keywords}
Galaxies: evolution; galaxies: elliptical and lenticular, cD; ISM: dust, extinction; quasars: general; cosmology: observations. 
\end{keywords}

\section{Introduction} 
In the last few years, IR surveys have revealed a 
large dust content in quasar (QSO) hosts at high redshift. 
Recent Atacama Large Millimeter/submillimeter Array (ALMA) observations 
of systems at redshift z$>$6 have shown the presence of large dust masses in QSO hosts 
(Wang et al 2010, 2011, 2013). 
Their large far-infrared luminosities indicate that vigorous starbursts are present in the 
centre of these systems, (Bertoldi et al. 2003), with star formation rates of 
the order of $\sim 10^3  M_{\odot}/yr$. 
Moreover, attempts have been carried on to constrain the dynamical 
properties of such systems; in a few objects, the observed velocity 
maps have revealed 
a rotating, gravitationally bound gas component, which made possible 
an estimate of the dynamical mass (Walter et al. 2004). \\
The considerable amounts of dust present in these objects is in contradiction 
with standard scenarios for interstellar dust formation. 
In fact, dust is expected to be present in little amounts at high 
redshift, if it is mostly produced in the cold atmospheres of evolved 
asymptotic giant branch (AGB) stars on timescales of a few Gyr (Matsuura et al. 2009). 
At these redshifts, the Universe is still young, and intermediate 
mass stars (i.e. those with masses $2 \lesssim m/M_\odot< 8$ ) are still on the main sequence, 
hence they did not have enough time to produce considerable amounts of dust grains. 
This suggests that other dust production 
sources should be at play to justify the large dust masses observed in high-redshift 
galaxies, such as core-collapse supernovae (Todini \& Ferrara 2001; Maiolino et al. 2006) or dust accretion 
in molecular clouds (Pipino et al. 2011). 
All the observational constraints on the dust content of 
high-redshift QSOs 
are of invaluable importance not only 
for constraining the relative importance of dust production mechanisms 
in the early Universe, 
but also for probing the star formation histories (SFHs) and the mass buildup of 
spheroidal galaxies, a fundamental benchmark for 
galaxy formation theories. \\
In this paper, we compare the available data of 
high-redshift QSO hosts with chemical evolution 
models reproducing the dust content of local elliptical galaxies and their 
downsizing behaviour, in order to improve our 
understanding of dust production and of the main 
parameters regulating star formation (SF) in the early Universe. 
Our primary goal is to understand, through this comparison, 
whether the intense starbursts galaxies hosting 
high-$z$ QSOs may be associated with the progenitors of local ellipticals. 
Another aim of this work is to assess whether 
at high redshift, the star formation conditions 
are similar to those observed in local starbursts, 
which are known to obey to the popular Schmidt (1959)-Kennicutt (1998a) relation, characterised by 
a marked correlation between the surface star formation rate (SFR) density and the molecular gas 
density. 
Our models can also be useful tools to 
test the reliability of the dynamical mass estimates from CO and CII emission, 
to understand the implications for the formation of such objects, 
for the duration of their starbursts, and finally, 
their abundance at high redshift and to explore the possibility to detect 
massive spheroids in the fields of present and future galaxy surveys. 
 
This paper is organized as follows. 
In Section 2, we present our observational dataset. We describe the observations from 
which the systems studied in this work are drawn, and we present the basic 
calibrations which allow us to estimate the basic observables. \\
In Section 3, we present our chemical evolution models. 
We present the basic chemical evolution equations, illustrating the 
role of a few parameters which will turn out to play a key role in this study. \\
In Section 4 we present our results. 
In Section 5, we discuss the role of systematics in the observational data set, 
the main implications of our results 
and provide predictions for the detectability of the systems associated to the descendants 
of high-$z$ QSO hosts. Finally, in Sect. 6 we draw our conclusions.

\section{The observational dataset} 
\label{sec_obs}
We searched the literature and compiled to the best of our knowledge a list of all QSOs at $z>5.7$ that have been observed in the far-IR.
Fifty-eight high-z QSOs were found, either discovered by the SDSS (e.g. Fan et al. 2006, Jiang et al. 2009), the CFHQS (Willott et al. 2007), 
the NDFWS (Cool et al. 2006) or the UKIDSS (Mortlock et al. 2009) survey.
The whole object list is shown in Table~1. Most of the far-IR observations were performed by the IRAM Max-Planck Millimeter Bolometer Array 
(MAMBO, Kreysa et al. 1998) at 250 GHz (1.2 mm); data from ALMA were also available for several objects. 
Seventeen objects were detected, whose far-IR luminosity $L_{FIR}$ (42.5-122.5 $\mu m$) was derived by fitting an 
optically thin grey-body to their far-IR spectral 
energy distribution (e.g. Wang et al. 2008). For most QSOs the flux density at 250 GHz ($S_{250}$) is the only far-IR datapoint available: 
in these cases, a dust temperature of $T_d=47$~K and emissivity index $\beta=1.6$ were assumed (e.g. Wang et al. 2008, 2010, 2011a, 2011b; Omont et al. 2013), 
which are consistent with the average values 
observed in QSO hosts at $z>2$ (Beelen et al. 2006). When $L_{FIR}$ was not directly available from the literature, we 
converted $S_{250}$ (or its 3$\sigma$ upper limit) into $L_{FIR}$ by using the same assumptions on the gray-body parameters as described above. 
This leads to the following conversion (Omont et al. 2013):
\begin{equation}
L_{FIR}(10^{12}L_{\odot}) = 2.3 \times S_{250}(mJy),
\end{equation}
for an object at $z=6.1$, which is also accurate to within 6\% (i.e. well below the statistical uncertainties) for the whole redshift range of the sample. 
The SFR and dust mass $M_d$ were derived directly from $L_{FIR}$ and are mostly quoted in the literature papers (see Table 1). 
Whenever no such values were reported, we derived them from $L_{FIR}$ as follows: we integrated a grey-body spectrum with $T_d=47$~K and $\beta=1.6$
to convert the far-IR (42.5-122.5$\mu$m) to total-IR (8-1000$\mu$m) luminosity ($L_{8-1000}\equiv L_{IR}=1.4\times L_{FIR}$), and then used the relation
$SFR(M_{\odot}/yr) = 1.74\times 10^{-10} \times L_{IR}(L_{\odot})$ (Kennicutt 1998b). The SFR values or their $3\sigma$ upper limits were then 
obtained from $L_{FIR}$ with the following conversion:
\begin{equation}
SFR(M_{\odot}/yr) = 243.6 \times L_{FIR}(10^{12}L_{\odot}).
\label{sfr_to_lir}
\end{equation}
The $L_{FIR}$-SFR conversion reported in Eq.~\ref{sfr_to_lir}  is 
generally considered valid in starbursts with ages less than $10^8$ yr. 
In older systems or in galaxies characterised by low-star formation efficiency,  
the conversion is 
uncertain due to the presence of evolved stars, 
contributing to dust heating, a different dust optical depth and the possible presence of an active galactic nucleus (Kennicutt 1998b). 
Another source of uncertainty is the assumed stellar initial mass function (IMF), which 
plays a major role in models of stellar population synthesis, where it is used as weight 
for the theoretical simple stellar populations, then summed together to synthesize the luminosities of galaxies 
(Kennicutt 1998b). 

As discussed later in Sect.~\ref{sect_imf}, in general, a top-heavy IMF, particularly rich in massive stars, 
gives place to a larger dust yield per stellar mass with respect to a standard Salpeter (1955) IMF;  
at fixed star formation rate, this corresponds to a larger $L_{FIR}$ to SFR ratio (see also Rengarajan \& Mayya 2004).

The total dust mass $M_d$ is obtained from the relation $M_d=L_{FIR}/\int 4\pi \kappa_\nu B_\nu(T) d\nu$, where $B_\nu(T)$ is the Planck function and 
$\kappa_\nu=\kappa_0(\nu/\nu_0)^\beta$ is the dust absorption coefficient, we again assumed $T_d=47$~K and $\beta=1.6$ and $\kappa_0=18.75$ cm$^2$g$^{-1}$
at 125$\mu m$ (Hildebrand 1983) as commonly used in the literature (Wang et al. 2011, Omont et al. 2013). 
This leads to the conversion:
\begin{equation}
M_d(10^8 M_{\odot}) = 0.56 \times L_{FIR}(10^{12}L_{\odot}).
\end{equation}
For these dust temperature values, typical of FIR and CO luminous quasars (Wang et al. 2011), no substantial effects of the cosmic 
microwave background is expected on the derived dust masses and, in case of ideal coupling between dust and gas 
temperatures, even on the derived molecular gas masses (Da Cunha et al. 2013). 

We then searched for any information about the molecular and atomic gas content of these systems, as traced by the CO emission line(s) or the
[CII]158$\mu$m emission line, respectively. Observations of CO lines were available for 17 objects (either from IRAM PdB, EVLA, ALMA). Twelve objects were detected, for which the molecular gas mass was derived by assuming $M_{H_2}=\alpha \times L'_{CO(1-0)}$, where $\alpha=0.8 M_{\odot} (K km\;s^{-1} pc^2)^{-1}$
is the CO intensity-to-gas mass conversion factor appropriate for ultra-luminous infrared galaxies (ULIRGs, Downes \& Solomon 1998) and $L'_{CO(1-0)}$ is the luminosity of the CO(1-0) line expressed as
the velocity-integrated brightness temperature times the source area (it has units of $K km\;s^{-1} pc^2)$. When considering the CO(1-0) line luminosity in
units of $L_{\odot}$, $L_{CO(1-0)}$ rather than $L'_{CO(1-0)}$ (which are linked by the relation $L_{CO(1-0)}=3\times10^{-11}\nu^3_{CO(1-0)}L'_{CO(1-0)}$), 
the relevant equation to convert CO luminosities into molecular gas masses becomes:
\begin{equation}
M_{H_2}(M_{\odot}) = 1.6 \times10^4\times L_{CO(1-0)}(L_{\odot})
\end{equation}

To get the luminosity of the CO(1-0) line from the higher-order CO transitions that are observed in these high-z QSOs, the line ratios observed for the 
z=6.43 QSO J1148+5251 are normally assumed (Riechers et al. 2009). Whenever the molecular gas mass is not reported directly from the literature papers, 
we derive it (or its 3$\sigma$ upper limit) from the CO line luminosity by using the same conversions as above. 

Observations of the [CII]158$\mu$m line are also available for 9 QSOs, either from IRAM PdB or ALMA observations; 8 objects have been detected.
In order to convert the [CII] line luminosity into an atomic gas mass, we followed 
Eq.1 of Hailey-Dunsheath et al. (2010) assuming a C$^+$ abundance 
per hydrogen atom of $1.4\times10^{-4}$, a gas temperature of T=200 K, and a gas density $n>>n_{crit}$, where $n_{crit}$ is the critical 
density of the [CII]158$\mu m$ transition ($\sim3\times10^3$~cm$^{-3}$). Under these assumptions (see also Maiolino et al. 2012) 
the following conversion holds:
\begin{equation}
M_{HI}(M_{\odot}) = 0.96 \times  L_{[CII]}(L_{\odot})
\label{eq_cii}
\end{equation}
All the $M_{HI}$ values in Table~1 have been derived by ourselves based on the $L_{[CII]}$ values reported in the literature. 
It is worth noting that there may be an uncertain fraction of [CII] not associated with the cold neutral medium, 
where the bulk of the atomic gas should belong. 
As oulined in Hailey-Dunsheath et al. (2010), the calibration reported in Eq. ~\ref{eq_cii} is valid for the 
atomic mass associated to dense photodissociation regions which account for at least 70\% of the 
[CII] emission in starbursts (e.g. Colbert et al. 1999). 

Finally, we recorded the measurements of the dynamical mass based either on the CO or [CII] line measurements, which have been obtained 
by the following relation (e.g. Wang et al. 2013):
\begin{equation}
M_{dyn}sin^2i (M_{\odot})= 6.5\times10^4 [FWHM(km/s)]^2\; D(kpc),
\end{equation}

where $i$ is the inclination angle between the gas disc and the line of sight and $D$ is the disc diameter. 
We used the conversion above to estimate $M_{dyn}sin^2i$ for three objects for which it was not reported (except for those showing evidence of 
complex, disturbed gas dynamics). Since typical disc sizes of a few kpc are now being observed by ALMA at sub-arcsec resolution 
(1"=5.7 kpc at $z=6$ for a concordance cosmology with $\Omega_m=0.3$, $\Omega_\Lambda=0.7$ and $h=0.7$), following previous works 
(e.g. Wang et al. 2010) we assumed a fiducial size of $D$=5 kpc for those objects observed at poor angular resolution 
(see Table 1); the effects of assuming a range of sizes will be discussed in Sect.~\ref{sect_uncert}. 

A measurement of $M_{dyn}sin^2i$ is available for 13 objects; for 7 objects we were able to collect an estimate of $M_d$, $M_H{_2}$, $M_{HI}$ and
$M_{dyn}sin^2i$. \\
For the sake of consistency, when discussing both data and model results, by {\it dynamical mass} we mean the total baryonic mass, i.e. the 
sum of the stellar mass and the gas mass. 
When comparing data with models, we use the dynamical mass estimates reported in Tab.~\ref{tab_obs} at 
face value, i.e. we assume that, in general, $sin^2i\sim 1$. 

For our observational sample, the stellar masses are defined as 
\begin{equation}
M_{star}=M_{dyn}-M_{gas}, 
\label{eq_mdyn}
\end{equation}
where $M_{gas}=M_{HI}+M_{H_2}$, i.e. we neglect the presence of dark matter.  
The contribution of supermassive black holes and dust to the dynamical 
mass is minor compared to those of gas and stars, thus it is neglected.

\begin{table*} 
\caption{Observational dataset of QSO hosts}
\begin{tabular}{lcccrrrrcrrrcc}
\hline \hline
Name& Survey& $m_{1450}$& $z$& R1& $L_{FIR}$& SFR& $M_{dust}$& R2& $M_{H2}$& $M_{HI}$& $M_{dyn}sin^2i$& R3\\
(1)& (2)& (3)& (4)& (5)& (6)& (7)& (8)& (9)& (10)& (11)& (12)& (13)\\
\hline
J0002+2550& S& 19.02& 5.80& 1&       $<4.2$&  	     $<$1023&  $<$2.4& 1&	$\cdots$&	 $\cdots$&		   $\cdots$& $\cdots$\\ 
J0005-0006& S& 20.23& 5.83& 1&       $<3.4$&  	      $<$828&  $<$1.9& 2&	$\cdots$& 	 $\cdots$&  		   $\cdots$& $\cdots$\\ 
J0033-0125& C& 21.78& 6.13& 2&  $2.6^{+0.8}_{-0.8}$&     633&     1.5& 2& 	$\cdots$&        $\cdots$&  		   $\cdots$& $\cdots$\\ 
J0050+3445& C& 19.84& 6.25& 3& 	     $<5.2$&  	     $<$1266&  $<$2.9& 3&	$\cdots$& 	 $\cdots$&  		   $\cdots$& $\cdots$\\ 
J0055+0146& C& 21.82& 6.02& 4& 	     $<3.2$&  	      $<$779&  $<$1.8& 3&	$\cdots$& 	 $\cdots$&  		   $\cdots$& $\cdots$\\ 
J0102-0218& C& 22.02& 5.95& 4& 	     $<2.6$&  	      $<$633&  $<$1.5& 3&	$\cdots$& 	 $\cdots$&  		   $\cdots$& $\cdots$\\ 
J0129-0035& S& 22.28& 5.78& 5&  $5.2^{+0.9}_{-0.9}$&  	1266&     2.9& 4& $1.2^{+0.2}_{-0.2}$&   $0.18^{+0.03}_{-0.03}$&       0.90& 1,2\\ 
J0136+0226& C& 22.04& 6.21& 3&	     $<6.7$&  	     $<$1632&  $<$3.8& 3&	$\cdots$& 	 $\cdots$&  	     	   $\cdots$& $\cdots$\\ 
J0203+0012& U& 20.97& 5.86& 6&  $4.4^{+1.1}_{-1.1}$&  	1072&     2.5& 4& 	$<1.0$&        	 $\cdots$&  	     	   $\cdots$& 1\\ 
J0210-0456$^a$& C& 22.57& 6.44& 7&  $0.3^{+0.1}_{-0.1}$&  	  48&    0.15& 5&	$<1.0$&   	 $0.03^{+0.01}_{-0.01}$&       0.65& 3,4\\ 
J0216-0455& C& 24.15& 6.01& 4&	     $<3.9$&  	      $<$950&  $<$2.2& 3&	$\cdots$& 	 $\cdots$&  		   $\cdots$& $\cdots$\\ 
J0221-0802& C& 21.98& 6.16& 3&	     $<9.4$&  	     $<$2290&  $<$5.3& 3&	$\cdots$& 	 $\cdots$&  		   $\cdots$& $\cdots$\\ 
J0227-0605& C& 21.41& 6.20& 4&	     $<3.7$&  	      $<$901&  $<$2.1& 3&	$\cdots$& 	 $\cdots$&  		   $\cdots$& $\cdots$\\ 
J0239-0045& S& 22.15& 5.82& 5&       $<3.9$&  	      $<$950&  $<$2.2& 4& 	$\cdots$&        $\cdots$&  		   $\cdots$& $\cdots$\\ 
J0303-0019& S& 21.28& 6.07& 8&	     $<3.5$&  	      $<$852&  $<$2.0& 2&	$\cdots$& 	 $\cdots$&  		   $\cdots$& $\cdots$\\ 
J0316-1340& C& 21.72& 5.99& 3&	     $<9.9$&  	     $<$2421&  $<$5.6& 3&	$\cdots$& 	 $\cdots$&  		   $\cdots$& $\cdots$\\ 
J0353+0104& S& 20.22& 6.05& 8&	     $<3.2$&  	      $<$779&  $<$1.8& 2&	$\cdots$& 	 $\cdots$&  		   $\cdots$& $\cdots$\\ 
J0836+0054& S& 18.81& 5.82& 9&	     $<6.6$&  	     $<$1603&  $<$3.7& 6& 	$<1.5$& 	 $\cdots$&  		   $\cdots$& 5\\ 
J0818+1722& S& 19.34& 6.00& 10&  $3.4^{+1.1}_{-1.1}$&  	 828&     1.9& 1&	$\cdots$& 	 $\cdots$&  		   $\cdots$& $\cdots$\\ 
J0840+5624$^{a,b}$& S& 20.04& 5.85& 10&  $7.6^{+1.5}_{-1.5}$&  	1851&     4.3& 1& $1.4^{+0.5}_{-0.5}$& 	 $\cdots$&  		       4.20& 3\\ 
J0841+2905& S& 19.61& 5.96& 11&      $<3.0$&  	      $<$731&  $<$1.6& 2& 	$\cdots$& 	 $\cdots$&  		   $\cdots$& $\cdots$\\ 
J0842+1218& S& 19.58& 6.08& 12&	     $<3.8$&  	      $<$925&  $<$2.2& 2&	$\cdots$& 	 $\cdots$&  		   $\cdots$& $\cdots$\\ 
J0927+2001$^a$& S& 19.87& 5.79& 10& $12.0^{+2.8}_{-2.8}$&  	2972&     6.9& 1& $3.0^{+0.4}_{-0.4}$& 	 $\cdots$&     	  	      20.90& 8,9\\ 
J1030+0524& S& 19.66& 6.28& 9&       $<7.6$&  	     $<$1863&  $<$4.3& 6& 	$\cdots$& 	 $\cdots$&  		   $\cdots$& $\cdots$\\ 
J1044-0125$^b$& S& 19.21& 5.80& 13&  $5.3^{+0.8}_{-0.8}$&  	1291&     3.0& 2& $0.7^{+0.1}_{-0.1}$&   $0.15^{+0.04}_{-0.04}$&       0.80& 2,6\\ 
J1048+4637$^b$& S& 19.25& 6.23& 14&  $6.9^{+0.9}_{-0.9}$&  	1681&     3.9& 7& $1.0^{+0.2}_{-0.2}$&   $\cdots$&      	       4.50& 6\\ 
J1059-0906& C& 20.75& 5.92& 3&	     $<5.7$&  	     $<$1389&  $<$3.2& 3&	$\cdots$& 	 $\cdots$&  		   $\cdots$& $\cdots$\\ 
J1120+0641$^c$& U& 20.40& 7.08& 15&  $1.4^{+0.4}_{-0.4}$&  	 346&     0.8& 8& 	$\cdots$&   	 $0.11^{+0.02}_{-0.02}$&       $<$3.60& 7\\ 
J1137+3549& S& 19.63& 6.01& 10&       $<5.2$&  	     $<$1267&  $<$2.9& 1&	$\cdots$&        $\cdots$&                 $\cdots$& $\cdots$\\ 
J1148+5251$^d$& S& 19.03& 6.43& 14& $12.0^{+1.4}_{-1.4}$&  	2923&     7.0& 7& $2.4^{+0.2}_{-0.2}$&   $0.40^{+0.03}_{-0.03}$&       4.60& 8,9\\ 
J1250+3130& S& 19.64& 6.13& 10&	     $<4.2$&  	     $<$1023&  $<$2.4& 1& 	$\cdots$&        $\cdots$&  		   $\cdots$& $\cdots$\\ 
J1306+0356& S& 19.55& 5.99& 9&	     $<6.9$&  	     $<$1698&  $<$3.9& 6&	$\cdots$& 	 $\cdots$&  		   $\cdots$& $\cdots$\\ 
J1319+0950& U& 19.65& 6.13& 16& $10.0^{+1.3}_{-1.3}$&  	2436&     5.7& 4& $1.5^{+0.3}_{-0.3}$&   $0.41^{+0.08}_{-0.08}$&       8.80& 2,6\\ 
J1335+3533& S& 19.89& 5.93& 10&  $5.5^{+1.2}_{-1.2}$&  	1340&     4.6& 1& $1.8^{+0.2}_{-0.2}$&   $\cdots$&      	       3.10$^b$& 6\\ 
J1411+1217& S& 19.97& 5.93& 1&       $<3.8$&  	      $<$926&  $<$2.1& 1& 	$\cdots$& 	 $\cdots$&   		   $\cdots$& $\cdots$\\ 
J1425+3254$^b$& N& 20.62& 5.85& 17&  $5.4^{+1.2}_{-1.2}$&  	1315&     3.0& 2& $2.0^{+0.4}_{-0.4}$& 	 $\cdots$&     		      15.60& 6\\ 
J1427+3312& N& 20.33& 6.12& 18&       $<4.6$&  	     $<$1121&  $<$2.6& 2& 	$\cdots$& 	 $\cdots$&  		   $\cdots$& $\cdots$\\ 
J1429+5447& C& 20.94& 6.21& 3&  $8.0^{+1.2}_{-1.2}$&    1949&     4.5& 3& $2.9^{+0.4}_{-0.4}$& 	 $\cdots$&  		   $\cdots$& 3\\ 
J1436+5007& S& 20.16& 5.83& 10&       $<5.4$&  	     $<$1315&  $<$3.1& 1& 	$\cdots$& 	 $\cdots$&  		   $\cdots$& $\cdots$\\ 
J1509-1749& C& 19.82& 6.12& 2&	     $<3.2$&  	      $<$780&  $<$1.8& 3&	$\cdots$& 	 $\cdots$&  		   $\cdots$& $\cdots$\\ 
J1602+4228& S& 19.86& 6.07& 1&	     $<5.8$&  	     $<$1413&  $<$3.3& 1&	$\cdots$& 	 $\cdots$&  		   $\cdots$& $\cdots$\\ 
\hline			    		       
\hline	
\end{tabular}
\label{tab_obs}
\end{table*}

\begin{table*} 
\contcaption{}
\begin{tabular}{lcccrrrrcrrrcc}
\hline \hline
Name& Survey& $m_{1450}$& $z$& R1& $L_{FIR}$& SFR& $M_{dust}$& R2& $M_{H2}$& $M_{HI}$& $M_{dyn}sin^2i$& R3\\
(1)& (2)& (3)& (4)& (5)& (6)& (7)& (8)& (9)& (10)& (11)& (12)& (13)\\
\hline
J1621+5155& S& 19.89& 5.71& 19&	     $<3.7$&  	      $<$901&  $<$2.1& 2&	$\cdots$& 	 $\cdots$&  		   $\cdots$& $\cdots$\\ 
J1623+3112& S& 20.13& 6.22& 1&	     $<3.6$&  	      $<$877&  $<$2.0& 1& 	$<1.6$& 	 $\cdots$&  		   $\cdots$& 3\\ 
J1630+4012& S& 20.64& 6.05& 14&	     $<4.2$&  	     $<$1023&  $<$2.4& 2& 	$<0.8$& 	 $\cdots$&  		   $\cdots$& 5\\ 
J1641+3755& C& 21.30& 6.04& 2&	     $<3.2$&  	      $<$780&  $<$1.8& 2& 	$\cdots$& 	 $\cdots$&  		   $\cdots$& $\cdots$\\ 
J2053+0047& S& 21.20& 5.92& 5&	     $<4.4$&  	     $<$1072&  $<$2.5& 4&	$\cdots$& 	 $\cdots$&  		   $\cdots$& $\cdots$\\ 
J2054-0005& S& 20.60& 6.06& 8&  $6.1^{+0.8}_{-0.8}$&  	1486&     3.1& 2& $1.2^{+0.2}_{-0.2}$&   $0.33^{+0.05}_{-0.05}$&       1.20& 2,6\\ 
J2100-1715& C& 21.37& 6.09& 3&	     $<4.1$&  	      $<$991&  $<$2.3& 3&	$\cdots$& 	 $\cdots$&  		   $\cdots$& $\cdots$\\ 
J2147+0107& S& 21.65& 5.81& 5&       $<4.4$&  	     $<$1072&  $<$2.5& 4& 	$\cdots$&        $\cdots$&  		   $\cdots$& $\cdots$\\ 
J2229+1457& C& 21.90& 6.15& 3&	     $<5.5$&  	     $<$1345&  $<$3.1& 3&	$\cdots$& 	 $\cdots$&  		   $\cdots$& $\cdots$\\ 
J2242+0334& C& 22.09& 5.88& 3&	     $<4.2$&  	     $<$1026&  $<$2.4& 3&	$\cdots$& 	 $\cdots$&  		   $\cdots$& $\cdots$\\ 
J2307+0031& S& 21.73& 5.87& 5&	     $<3.8$&  	      $<$926&  $<$2.1& 4&	$\cdots$& 	 $\cdots$&  		   $\cdots$& $\cdots$\\ 
J2310+1855& S& 19.30& 6.00& 20& $17.2^{+1.0}_{-1.0}$&   4190&     9.7& 9& $5.1^{+0.4}_{-0.4}$&   $0.16^{+0.04}_{-0.04}$&       4.90& 2\\ 
J2315-0023& S& 21.34& 6.12& 8&	     $<4.1$&  	      $<$999&  $<$2.3& 2&	$\cdots$& 	 $\cdots$&  		   $\cdots$& $\cdots$\\ 
J2318-0246& C& 21.55& 6.05& 4&	     $<3.9$&  	      $<$942&  $<$2.2& 3&	$\cdots$& 	 $\cdots$&  		   $\cdots$& $\cdots$\\ 
J2329-0301& C& 21.65& 6.43& 2&	     $<0.2$&  	       $<$40&  $<$0.1& 5&	$\cdots$&         $<0.01$&  		   $\cdots$& 4\\ 
J2329-0403& C& 21.96& 5.90& 4&	     $<4.3$&  	     $<$1059&  $<$2.4& 3&	$\cdots$& 	 $\cdots$&  		   $\cdots$& $\cdots$\\ 
J2356+0023& S& 21.77& 6.00& 5&	     $<2.8$&  	      $<$682&  $<$1.6& 4&	$\cdots$& 	 $\cdots$&  		   $\cdots$& $\cdots$\\ 
\hline			    		       
\hline		
\end{tabular}
\begin{flushleft}
Column description: (1) QSO name. Objects are sorted by increasing RA. (2) Discovery survey: S=SDSS; C=CFHQS; U=UKIDSS; N=NDWFS. 
(3) Apparent AB magnitude at 1450\AA \, rest-frame. (4) Redshift. (5) Reference to QSO discovery paper and quantities reported in Columns (1)-(4): 
1= Fan et al. (2004); 
2= Willott et al. (2007); 3=Willott et al. (2010a); 4=Willott et al. (2009); 5=Jiang et al. (2009); 6=Venemans et al. (2007); 7=Willott et al. (2010b);
8=Jiang et al. (2008); 9=Fan et al. (2001); 10=Fan et al. (2006); 11= Goto (2006); 12=De Rosa et al. (2011); 13=Fan et al. (2000); 14=Fan et al. (2003);
15= Mortlock et al. (2011); 16=Mortlock et al. (2009); 17=Cool et al. (2006); 18=McGreer et al. (2006); 19= Wang et al. (2008); 
(6) Far-IR luminosity in units of $10^{12}L_{\odot}$. (7) Star formation rate derived from $L_{FIR}$ 
in units of $M_{\odot}$yr$^{-1}$. (8) Dust mass derived from $L_{FIR}$ in units of $10^{8}M_{\odot}$. (9) Reference to $L_{FIR}$ and related quantities 
in Columns (6)-(8): 1=Wang et al. (2007); 2=Wang et al. (2008); 3=Omont et al. (2013); 4=Wang et al. (2011a); 5=Willott et al. (2013); 6=Petric et al. (2003);
7=Bertoldi et al. (2003); 8=Venemans et al. (2012); 
 (10) Molecular gas mass derived from CO line measurements
in units of $10^{10}M_{\odot}$. (11) Atomic gas mass derived from CII line measurements in units of $10^{10}M_{\odot}$. 
(12) Dynamical mass derived either from CO or CII line measurements in units of $10^{10}M_{\odot}$. (13) Reference to molecular and/or atomic gas measurements
in Columns (10)-(12): 1=Wang et al. (2011a); 2=Wang et al. (2013); 3=Wang et al. (2011b); 4=Willott et al. (2013); 5=Maiolino et al. (2007); 
6=Wang et al. (2010); 7=Venemans et al. (2012); 8=Walter et al. (2003); 9=Carilli et al. 2007. Upper limits are $3\sigma$. See text for details.	

Notes:$^a$the value of $M_{dyn}sin^2i$ is derived in this work using Eq.6. $^b$Disc diameter (see Eq.6) assumed to be D=5 kpc, either 
here or in the original work. $^c$Disc diameter assumed to be $<$10 kpc as in the original work. For all the remaining systems, the disc size as measured 
in the original paper is used. $^d$A broad [CII] component, likely produced by outflowing gas, is found in this object (Maiolino et al. 2012). 
The mass of the atomic hydrogen reported here is derived from the narrow [CII] component as measured by Walter et al. (2009).

Most measurements of $L_{FIR}$, SFR and $M_d$ are based on a single datapoint at 250 GHz from MAMBO, and therefore heavily depend on our
assumed FIR SED with $T_d=47$ K and $\beta$=1.6.  For a given flux density at 250 GHz, the values of $L_{FIR}$ (and SFR) would be $\sim3$ times lower 
if $T_d=33$ K is assumed instead of $T_d=47$ K. On the contrary, $M_{dust}$ would be $\sim2.4$ times larger (Omont et al. 2013). We therefore remark
that significant systematic uncertainties can affect the measurements reported in this Table. A full discussion of systematic uncertainties is presented
in Section 5.1\\

\end{flushleft}
\end{table*}

\section{Chemical evolution models for elliptical galaxies}
\label{sec_mod}
In this section, we present a brief description of the chemical
evolution models used in this work. 
The models are calibrated in order to reproduce various features of 
elliptical galaxies, including their local spectral energy distributions (SEDs) and those of their high-redshift counterparts 
(Schurer et al. 2009; Fan et al. 2013). 
More detailed descriptions can be found in Calura et al. (2008) and Pipino et al. (2011). 
In our scheme, elliptical galaxies form from the rapid collapse of 
a gas cloud with primordial chemical composition. 
The collapse is assumed to be described by an exponential infall law, with e-folding time $\tau_{inf}$. 
After the initial collapse, the galaxy is allowed to evolve as an 'open box' 
into the potential well of a dark matter halo. 
The rapid collapse triggers an intense and rapid star formation process, 
which lasts until a galactic wind, 
powered by the thermal energy injected by stellar winds and SNe (Ia, II) explosions, occurs. 
At that time, the gas thermal energy equates the gas binding energy and all the residual 
interstellar medium is assumed to be lost. After that time, the galaxies evolve passively. 

In the interstellar medium, the evolution of the fractional mass of the element $i$,   
$G_{i}$, is described by:

\begin{equation}
\label{eq_chem}
\dot{G_{i}}=-\psi(t)X_{i}(t) + R_{i}(t) + (\dot{G_{i}})_{inf} -
(\dot{G_{i}})_{out}
\end{equation}

where $G_{i}(t)=M_{gas}(t)X_{i}(t)/M_{lum}$ is the gas mass in 
the form of an element $i$ normalized to the total luminous mass 
$M_{lum}$ and $G(t)= M_{gas}(t)/M_{lum}$ is the fractional 
mass of gas present in the galaxy at the time $t$. 

The quantity $X_{i}(t)=G_{i}(t)/G(t)$ represents the 
abundance by mass of an element $i$, with
the summation over all elements in the gas mixture being equal 
to unity. $\psi(t)$ is the star formation rate, i.e. the  
amount of gas turning into stars per unit time. 
$R_{i}(t)$ represents the returned fraction of matter in the 
form of an element $i$ that the stars eject into the ISM through 
stellar winds and supernova explosions. 
This term includes the contribution from single low-and intermediate-mass stars, 
characterised by initial masses $m<8 M_{\odot}$, 
from type II SNe, originating from the explosion of massive stars with initial mass of 
$m>8 M_{\odot}$, and from type Ia SNe, for which we assume the single-degenerate (SD) scenario as proposed by Whelan \& Iben (1973). 
In this scenario, a C-O white dwarf in a binary system 
accretes mass from a non-degenerate companion until it reaches the Chandrasekhar mass ($\sim1.4M_{\odot}$) 
and explodes via C-deflagration, leaving no remnant. 
The fraction of stars in binary systems able to originate type Ia SNe is fixed by the present-day type IA SN rate 
observed in local galaxies (Matteucci \& Recchi 2001; Calura \& Matteucci 2006).

In Eq.~\ref{eq_chem}, the two terms 
$(\dot{G_{i}})_{inf}$ and  $(\dot{G_{i}})_{out}$ account for 
the infall of external gas and for galactic winds, respectively.
For the infall, we assume an exponential law 
\begin{equation}
(\dot{G_{i}})_{inf}=X_{i,inf} \exp{-t/\tau_{inf}}, 
\end{equation}
where $X_{i,inf}$ describes the chemical composition of the infalling gas, assumed to be primordial. 
The quantity $\tau_{inf}$ determines the timescale of the collapse; the values adopted for this quantity 
are reported in Tab.~\ref{tab_models}. 

The SFR (expressed in $M_{\odot}$/yr) is calculated as:
\begin{equation}
\psi(t) = \nu M_{gas}(t), 
\end{equation}
i.e. it is assumed to be proportional to the gas mass via a constant $\nu$, according to the Schmidt (1959) law. 
The star formation efficiency $\nu$ is allowed to vary as a function of the mass (see Tab.~\ref{tab_models}), 
in order to reproduce the “inverse wind model” of Matteucci (1994), an earlier version of the ``downsizing'' 
behaviour of galaxies. 

Another fundamental quantity is the stellar initial mass function (IMF), i.e. the mass distribution of 
stars born in a stellar population.  
For this quantity, 
in standard conditions,  we assume a simple power law $\phi(m)\propto m^{-1.35}$ (Salpeter 1955). \\
We will also investigate the effect of a Larson (1998) IMF, richer in massive stars 
and of the form
\begin{equation}
\phi_{L}(m)\propto m^{-1.35} \exp(-m_{c}/m), 
\label{eq_lar}
\end{equation}
i.e. a power law modulated with an 
exponential function.

\begin{figure*}
\epsfig{file=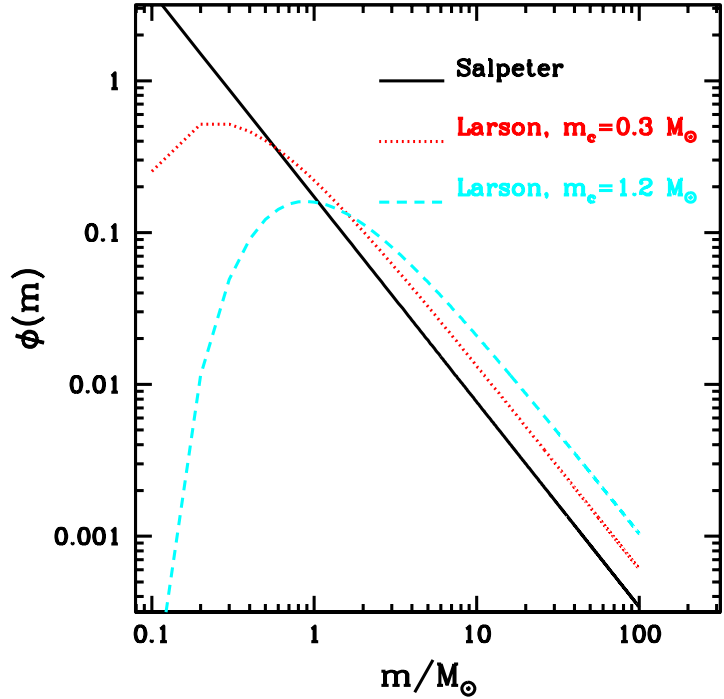,height=10cm,width=10cm}
\caption{Initial mass functions used for our chemical evolution models. 
The black solid line is a Salpeter (1955) IMF. The red dotted line and the cyan 
dashed line are Larson IMFs expressed by Eq.\ref{eq_lar}, computed assuming for the 
characteristic mass $m_{c}$ the values $0.3 M_{\odot}$, of the order of the local Jeans mass (Bate \& Bonnell 2005),  and $1.2 M_{\odot}$, respectively.} 
\label{imf}
\end{figure*}

The Salpeter (1955) and Larson (1998) IMF are plotted in Fig.~\ref{imf}; 
the presence of a characteristic mass in the latter is the main feature which distinguishes  
the two IMFs. In Fig.~\ref{imf}, we show a Larson IMF calculated with characteristic masses 
$m_c=0.3$ M$_\odot$ and $m_c=1.2$ M$_\odot$. 
The former value is of the order of the Jeans mass 
in local star-forming environments (Bate \& Bonnell 2005), whereas the latter value, which implies 
a top-heavy IMF, i.e. richer in massive stars, will be used later to study the properties of high-$z$ QSO hosts. 

In our chemical evolution models, stellar masses are calculated self-consistlently 
by taking into account the star formation history and the mass return from ageing stellar populations. 

The main features of the models used in this paper are summarised in Tab.~\ref{tab_models}. 
In the first column, the name of the model is shown. 
The second column shows the adopted total baryonic mass of the models.
The third, the fourth and the fifth columns indicate for each model, the effective radius, 
the star formation efficiency and the infall timescale, respectively.  
Finally, the sixth and seventh columns show the adopted stellar IMF and the dust accretion timescale, respectively. 
The first three models (M3E10, M1E11 and M1E12) are characterised by standard choices of the 
basic star formation and dust production parameters and by a standard  Salpeter IMF. 
The models M3E11S, M3E11D and M3E11L present instead non-standard assumptions for 
the main parameters, such as an enhanced star formation efficiency (M3E11S) 
combined with an increased dust accretion efficiency (M3E11D) and the additional 
choice of a Larson (1998) IMF (M3E11L).\\

\begin{table*}
\begin{flushleft}
\vspace{0cm}
\caption[]{Main parameters assumed for our chemical evolution models for elliptical galaxies.} 
\begin{tabular}{l|llllll}
\noalign{\smallskip}
\hline
\hline
\noalign{\smallskip}
 Model name            &   $M_{lum}$      &  $R_{eff}$     & $\nu$          & $\tau_{inf}$    &   IMF      &   $\tau_{0}$   \\      
                       &  $(M_{\odot})$   &   (kpc)       &  (Gyr$^{-1}$)   &  Gyr           &              &   (Gyr)        \\    
\noalign{\smallskip}                                                                                                                                         
\noalign{\smallskip}                                                                                                                                         
\hline                                                                                                                                                       
\noalign{\smallskip}                                                                                                                                         
M3E10                  & $3 \times 10^{10}$  & 2         &   5             &    0.5        &   Salpeter   & 0.01  \\              
M1E11                  & $1 \times 10^{11}$  & 3         &   10            &    0.4        &   Salpeter   & 0.01  \\              
M1E12                  & $1 \times 10^{12}$  & 10         &  20            &    0.2        &   Salpeter   & 0.01  \\              
\hline                                                                                                                                       
M3E11S                 & $3 \times 10^{11}$  & 3         &  60            &    0.4         &   Salpeter   & 0.01  \\              
M3E11D                 & $3 \times 10^{11}$  & 3         &  60            &    0.4         &   Salpeter   & 0.002  \\             
M3E11L                 & $3 \times 10^{11}$  & 3         &  60            &    0.4         &   Larson   & 0.002  \\               
\hline
\hline
\end{tabular}
\label{tab_models}
\end{flushleft}
\begin{flushleft}
Column description: (1) name of the model. (2)  adopted total baryonic mass. (3) Effective radius. 
(4) Star formation efficiency. (5) Infall timescale. (6) Initial mass function. (7) Dust accretion timescale. 
\end{flushleft}
\end{table*}

\subsection{Chemical evolution of interstellar dust}
\label{sec_mod_dust}
We adopt the formalism of Dwek (1998) to describe the chemical evolution of 
dust grains in the interstellar medium (ISM). 
More detailed descriptions of this formalism can be found in Calura et al. (2008) and Pipino 
et al. (2011). \\
In our models, we assume that only the main refractory elements, C, O, Mg, Si, S, Ca, Fe, are depleted into dust. 
Dust grains are divided into silicates, which include the elements O, Mg, Si, S, Ca and Fe, 
and carbon dust, which incorporates only C. 
Dust grains are first produced by stars and restored into the ISM, in which 
they can undergo other processes. 
Stellar dust grains can be produced mostly by two sources: \\
{\it{(i)}} in low and intermediate mass stars, in a cold shell during the AGB phase 
(Ferrarotti \& Gail 2006); the composition of dust grains in determined by the $C/O$ ratio in the 
cold envelope, so that if $C/O>1$, carbon dust (or graphite) is produced, whereas if $C/O<1$ only silicates are produced. \\
According to the prescriptions of Dwek (1998) used here, 
a 2 $M_{\odot}$ star produces up to $0.01 M_{\odot}$ of C dust, consistent with other estimates from the literature 
(Dwek \& Cherchneff 2011). 
{\it{(ii)}} Type II SNe, assumed to originate from the core-collapse of single stars of initial 
mass $m>8 M_{\odot}$. 
In these systems, dust grains are ejected into the ISM in fixed fractions of 
the refractory elements produced and restored (Dwek 1998). 
The prescriptions assumed for dust creation in type II SNe are the same as described in Pipino et al. (2011): 
a typical $20 M_{\odot}$ star releases 0.08  $M_{\odot}$ of dust, compatible with observational constraints from 
local SN remnants, providing dust masses between 0.01 $M_{\odot}$ and 0.1 $M_{\odot}$ 
(Rho et al. 2009; Barlow et al. 2010; Temim et al. 2012). \\
In this paper, we assume that no dust is produced in type Ia SNe. This assumption is supported by model 
results of nucleation and grain growth  in SN remnants, showing that the grains formed in the ejecta are 
completely destroyed in the hot gas swept up by the SN shocks (Nozawa et al. 2011), as well as the lack of 
cold dust in Ia SN remnants as shown by mid-far infrared observations with {\it Herschel} (Gomez et al. 2012).\\

Another mechanism for dust production is the accretion, or growth process, 
which may occur in dense molecular clouds. 
In our model, we assume that dust grains can undergo accretion 
on a typical timescale 
\begin{equation}
\tau_{growth} =  \frac{\tau_{0}}{1-G_{dust,i}/G_{i}}, 
\label{eq_gro}
\end{equation}
where $G_{dust,i}/G_{i}$ is the ratio between the mass fraction of the element $i$ into dust and 
its interstellar mass fraction. 
In this process, refractory elements can condensate onto pre-existing grain cores. 
In this work, we assume that grain growth occurs only during the starburst phase. 
The parameter $\tau_{0}$ corresponds to the typical accretion timescale. 
This quantity is a complex function of the interstellar metallicity, 
its density, temperature and of the grain size (Dwek 1998; Kuo \& Hirashita 2012). \\
As default assumption, here we adopt $\tau_{0}=0.01$ Gyr, as assumed also 
in Pipino et al. (2011), a value which is shorter than (but of the order of) the typical local molecular cloud 
lifetime in nearby galaxies (20-30 Myr, Fukui \& Kawamorua 2010). 
However, during the starburst phase, the conditions for star formation in proto-spheroids could be 
considerably different than in local SF regions of disc galaxies. 
The role of this parameter 
in determining the amount of dust in QSO hosts and its implications will be studied later in Sect.~\ref{sec_dust}. 

In this paper, we assume that a negligible amount of dust is produced in the broad line regions of the QSO (Elvis et al. 2002).
In fact, as shown by  Pipino et al. (2011), 
the cumulative dust mass produced by the QSO is three orders of magnitude lower than the cumulative effect 
of the dust growth, which is the major source of dust production in our model. 

In our model, dust grains 
can also be destroyed by SN shocks, on a timescale $\tau_{destr}= \xi^{-1} M_{gas}/R_{SN}$, 
where $R_{SN}$ represents the total type Ia+II SN rate (Dwek 1998; Calura et al. 2008). 
The quantity $\xi$ depends on the mass $M_{SNR}$ of the ISM swept by the SN remnant 
and on an average efficiency $\epsilon$, which assumes the value  $\epsilon=0.2$ 
in a three-phase medium, which consists of 
a cold neutral medium, a warm interstellar medium and a hot medium (McKee 1989).  
As suggested by McKee (1989), we thus assume 
\begin{equation}
\xi=\epsilon M_{SNR} = 1360 M_{\odot}.
\end{equation}

\begin{figure*}
\epsfig{file=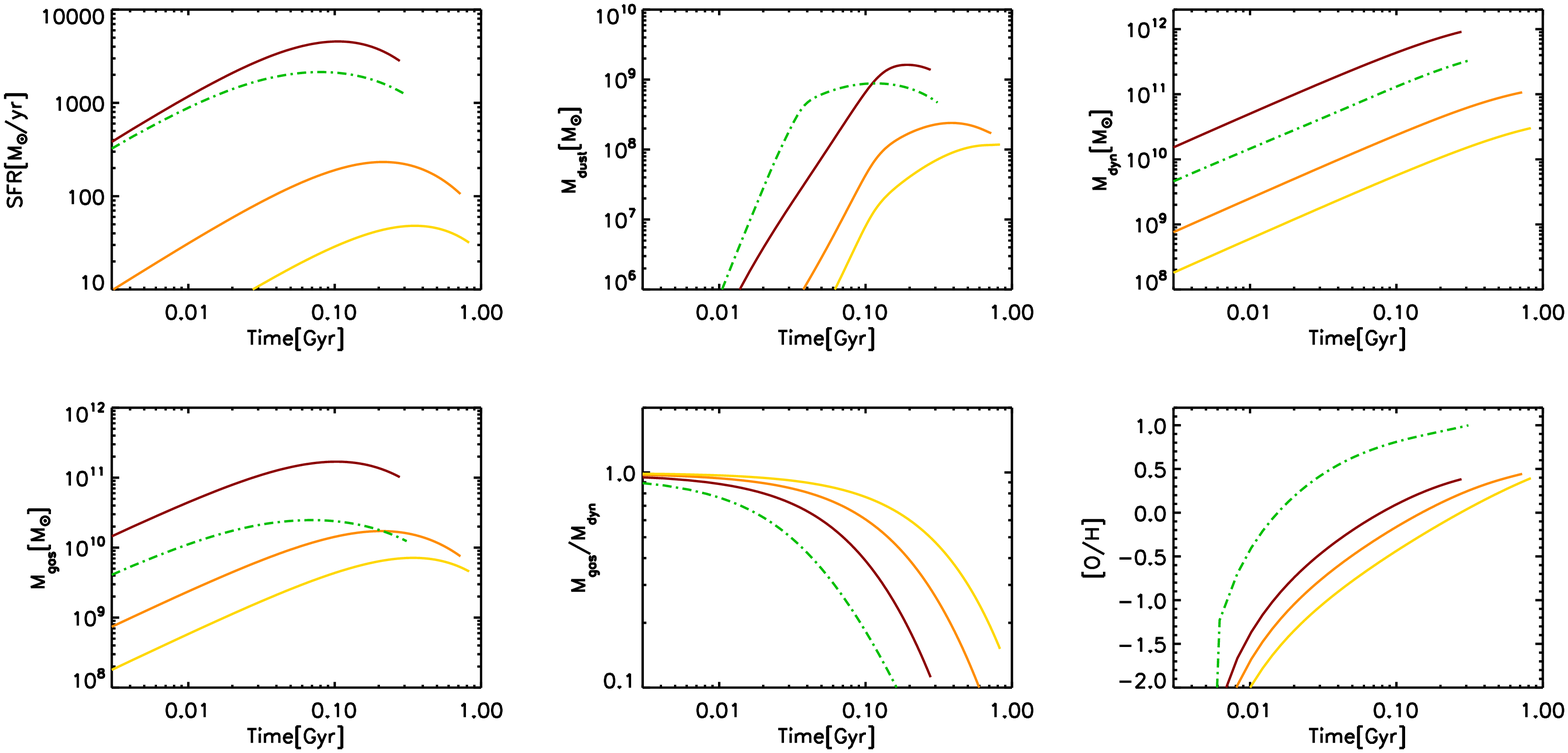,height=10cm,width=18cm}
\caption{Clockwise, from top-left: star formation history, evolution of the dust mass, of the dynamical mass, 
of the interstellar metallicty (as traced by the [O/H] ratio), 
of the gas-to-dynamical mass ratio and of the gas mass as a function of time for our  chemical evolution models
described in Tab.~\ref{tab_models}. In each panel, the red solid line, the orange solid line and the yellow solid line 
represents the model M1E12, M1E11 and M3E10, respectively. The green dot-dashed line represents our model M3E11L, 
which includes an increased star formation efficiency and a Larson (1998) IMF with $m_c=1.2$ M$_\odot$ (see eq.~\ref{eq_lar}).}
\label{fig_models}
\end{figure*}

In Fig. ~\ref{fig_models} we show, in clockwise sense starting from the top-left panel, 
the star formation histories, the evolution of the dust mass, of 
the dynamical mass, the gas-to-dynamical mass ratio, the gas mass evolution and the 
metallicity (here traced by [O/H]\footnote{In Fig.~\ref{fig_models}, the quantity  [O/H] is defined as 
\begin{equation}
[O/H]=log(O/H)-log(O/H)_{\odot},
\end{equation}
where $(O/H)$ and $(O/H)_{\odot}$  represent the interstellar and solar fraction of O with respect to H, respectively. 
Solar abundances are from Grevesse et al. (2010).}) 
as a function of 
time for the most representative models during the starburst phase. 
In the standard models (M3E10, M1E11 and M1E12), the downsizing is visible from the more extended   
SFHs in the less massive systems. 
The star formation is assumed to stop as soon as the conditions for the onset of the galactic wind 
are met, i.e. when the thermal energy of the ISM balances its binding energy. 

The model M3E11L assumes a higher star formation efficiency, a higher dust accretion efficiency 
and a top-heavy IMF, which 
results in the shortest starburst duration, 
in the quickest buildup of the dust mass and in the steepest decline of the gas-to-mass ratio. 

In each model, the dynamical mass $M_{dyn}$ grows with cosmic time to reach a value corresponding 
to the one adopted for the total baryonic mass $M_{lum}$, reported in Tab. ~\ref{tab_models}. 
The growth is driven mainly by the e-folding time of the 
exponential infall law, which regulates the amount of gas available for star formation. 
More massive systems have smaller infall timescales, as shown in Tab.~\ref{tab_models}, hence they experience a steeper 
growth of  $M_{dyn}$ as a function of time. 
 
A striking feature of Fig.~\ref{fig_models} is the evolution of the interstellar 
metallicity of model M3E11L which,  after $\sim 0.3$ Gyr, is more than 0.5 dex higher than that 
of the models assuming a Salpeter IMF. 
In principle, the measure of the interstellar metallicity in star-forming objects, 
already detectable through deep near-IR spectroscopy in systems at $z>3.5$ (Maiolino et al. 2008), 
offers a unique way to test the initial mass function of high-redshift galaxies.

\begin{figure*}
\epsfig{file=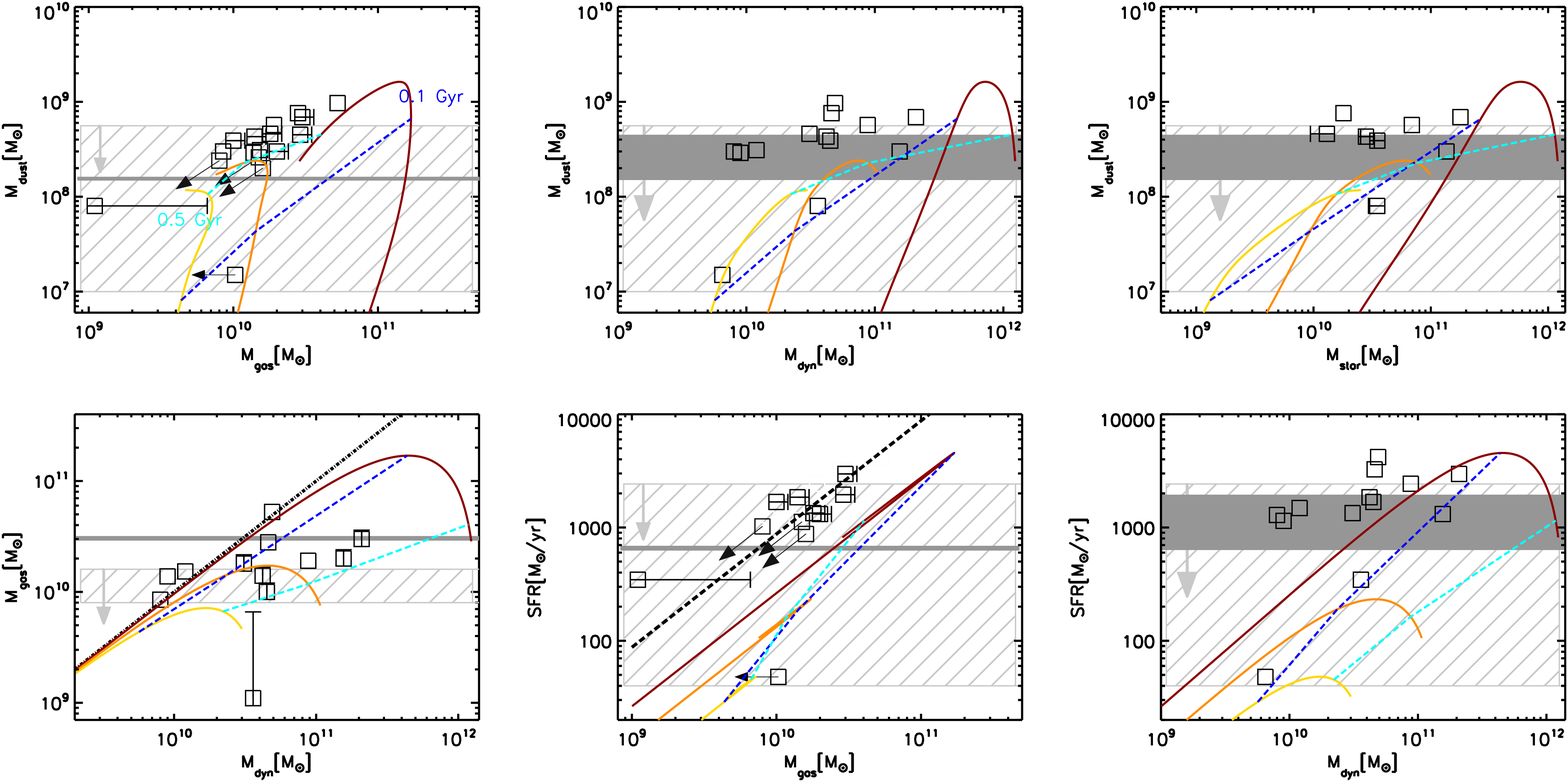,height=10cm,width=18cm}
\caption{Clockwise, from top-left: dust mass versus gas mass, dust mass vs dynamical mass, 
dust mass vs stellar mass, gas mass vs dynamical mass, SFR vs gas mass and SFR vs 
dynamical mass as observed in the sample of high-redshift QSO hosts and as predicted by chemical 
evolution models for elliptical galaxies. In Each panel, the light-gray hatched areas with 
an inscribed arrow enclose the systems which come without any information on the quantity plotted on the {\it x}-axis 
label, and with an upper limit for the {\it y}-axis quantity. 
The dark-gray shaded areas enclose the the systems without any information on the quantity plotted on the {\it x}-axis 
label, and with an observation for the {\it y}-axis quantity. The open squares represent the observations, 
plotted with an arrow if one of the {\it x-} or {\it y}-axis quantity represent upper limits, 
or with an error bar if only either their HI mass or only their H2 mass have been detected. 
If only the H$_{I}$ mass is available, the error bar extends to a H mass value  
corresponding to 6 times the H$_{I}$ mass, representing to the median $M_H/M_{H_{I}}$.  
In the plots including the dynamical mass, we remind that what we really plot is 
$M_{dyn/sin i^2}$. The dot-dashed straight line in the $M_{gas}$-$M_{dyn}$ plot is where any system with 
$M_{gas}$=$M_{dyn}$ would lie. 
The long-dashed straight line in the SFR-$M_{gas}$ plot represents a 
model with a star formation efficiency $\nu=60$ Gyr$^{-1}$. 
The dark red, orange, and yellow solid lines represent our results for the 
M1E12, M1E11 and M3E10 model, respectively. 
Finally, the blue and cyan short-dashed lines represent model isochrones at 0.1 and 0.5 Gyr, respectively. } 
\label{fig_dust}
\end{figure*}

\begin{figure*}
\epsfig{file=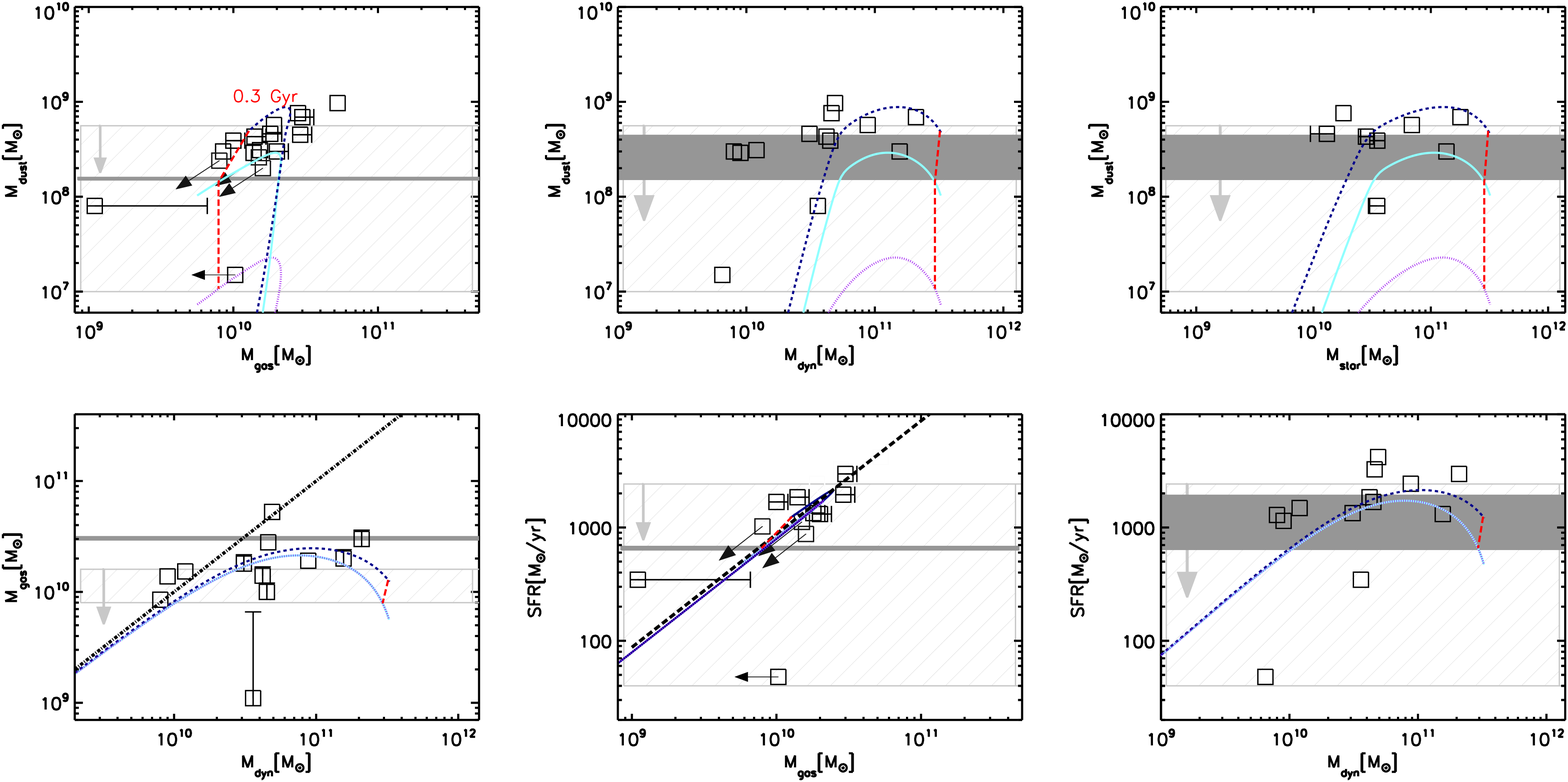,height=10cm,width=18cm}
\caption{Same as in Fig.~\ref{fig_dust}, but with the results of the models 
M3E11S (purple dotted lines), M3E11D (light-blue solid lines) and M3E11L (dark-blue dashed lines). 
The red long-dashed lines represent model isochrones at 0.3 Gyr. 
The model M3E11L includes an enhanced star formation efficiency, enhanced dust production 
and a Larson (1998) top-heavy IMF and can account for most of the observables considered in this figure.} 
\label{fig_dust_mod}
\end{figure*}
\begin{figure*}
\epsfig{file=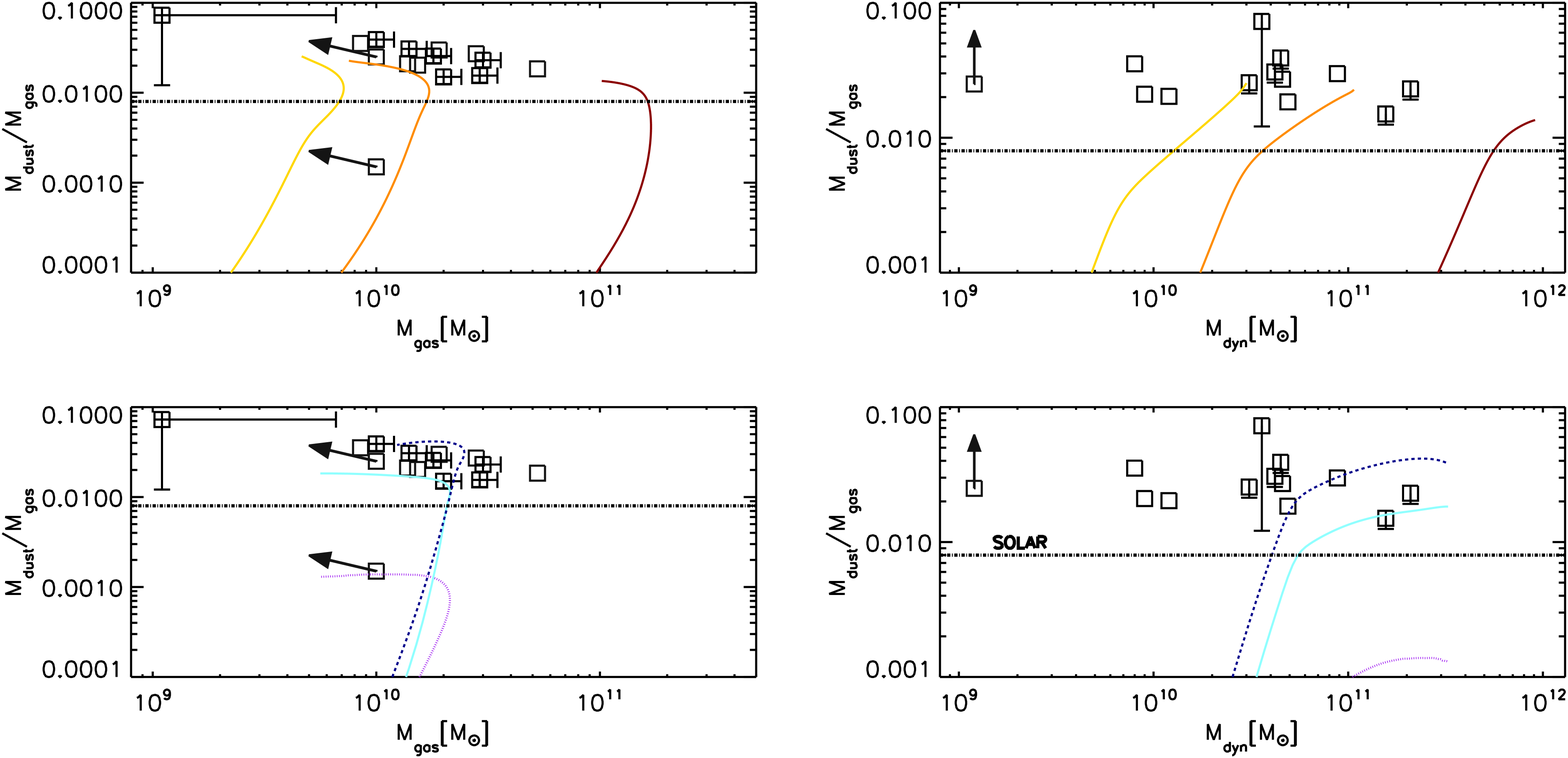,height=10cm,width=18cm}
\caption{Top-left panel: dust-to-gas ratio versus gas mass as observed in our dataset (open squares) and 
as calculated with our models. From left to right, the three solid curves represent the results of our models 
M3E10, M1E11 and M1E12. Top-right panel: dust-to-gas ratio versus dynamical mass; symbols and curves as above. 
The bottom-left and bottom-right panels show the dust-to-gas ratio versus gas mass and versus 
dynamical mass, respectively, as observed and as calculated with our models M3E11S (purple dotted line), 
M3E11D (cyan solid line), and M3E11L (dashed blue line). 
In each panel, the dot-dashed straight line represents a determination of the dust-to-gas ratio 
in the local Milky Way disc (Issa et al. 1990). }
\label{dtg}
\end{figure*}
\section{Results} 
\label{res}
In Fig. ~\ref{fig_dust}, we show various plots involving the observables presented in Tab.~\ref{tab_obs}. 
Starting from the top-left panel of Fig.~\ref{fig_dust}, in clockwise 
sense, the observed relation beween the dust mass and the gas mass, 
dust mass and dynamical mass, dust mass and stellar mass, 
SFR and dynamical mass, SFR and gas mass, and gas mass and dynamical mass for the 
set of QSO hosts are shown, compared with results from the chemical evolution models 
presented in Sect.~\ref{sec_mod}. 
 
In each plot, the data are divided in three groups: {\it (i)} the data 
with no information for the quantity on the {\it x}-axis and 
with an upper limit for the quantity on the {\it y}-axis (light-grey hatched regions); 
{\it (ii)} the data 
with no observation for the quantity on the {\it x}-axis and 
with a detection for the quantity on the {\it y}-axis (dark-grey shaded areas), 
and {\it (iii)} the data with both quantities available on the  {\it x}- and  {\it y}- axis 
(open squares), bearing an arrow in the case of upper limits or an error bar 

We assume $M_{gas}=M_{HI}+M_{H2}$ and, when either the H$_{I}$ or the H$_{2}$ mass is 
not available, we assume $M_{H2}/M_{HI}\sim 5$, corresponding to the average 
value in those systems where both quantities have been measured. 

The three solid lines are the results calculated for our models of elliptical galaxies. 
In these models, no special assumptions are made regarding the stellar IMF or dust production, i.e. 
a Salpeter IMF is assumed and the prescriptions regarding dust production are those described in Pipino et al. (2011). 
In each panel of Fig.~\ref{fig_dust}, the dashed lines crossing the solid curves  
represent isochrones at 0.1 (blue) and 0.5 (cyan) Gyr. 

As expected, the curves for the standard models are contained within 
the grey shaded areas, with the exception of the model for the most massive galaxy. 
For this model, the gas mass and dust mass values appear larger than the ranges enclosed within 
the shaded boxes. 

As far as the dynamical mass is concerned, we remind the reader that in Fig.~\ref{fig_dust} and in the following, 
what we really plots is $M_{dyn}/sin^2i$. As discussed in Sect.~\ref{sect_uncert}, for some systems 
this may represent an underestimate of their true dynamical mass.

As far as the observations are concerned (open squares in Fig.~\ref{fig_dust}), 
we note that the intermediate-mass  model (M1E11) encompasses  
a gas mass and dynamical mass range which include all the observed data. 
However, both the predicted dust masses and SFRs are lower with respect to the observed range.
To quantify this discrepancy, we can compare the mean observed dust mass for 
the systems with observations,  
$<M_{d}>=4 \cdot 10^{8} M_{\odot}$, 
with the maximum dust mass of the M1E11 model, $<M_{d,max}>=2.4 \cdot 10^{8} M_{\odot}$. 

The discrepancy between the maximum SFR value characterising model M1E11 and 
the observations (excluding the upper limits) is even larger, by a factor of 5. 

On the other hand, both  the dust masses and the SFRs obtained for the most massive model (M1E12) 
are compatible with the values 
of the detections, however the correct dust masses and SFRs are reproduced at gas and dynamical masses higher than 
the observations. 
This feature is visible for instance in the $M_{dust}-M_{gas}$, SFR-$M_{dyn}$, and $M_{dust}-M_{star}$ plots. 

In the $M_{gas}-M_{dyn}$ plot, for most of the observed systems the agreement between model results and observations is acceptable. 

The SFR-$M_{gas}$ plot is a probe of the validity of the  Schmidt-Kennicutt relations, which 
links SFR and H mass. In fact, it is easy to show that the common model assumption 
for the dependence of the SFR on the gas mass 
\begin{equation}
\psi(t) \propto  \nu M_{gas}(t)  
\end{equation}
implies $\psi(t) \propto \rho^{N}$, where $\rho$ is the gas density on galactic scale,
and with $N\sim 1.5$ 
(after having assumed a constant scale height for any galaxy), 
as shown empirically by Kennicutt (1998) in local star forming galaxies. 
It is important to stress that the SFRs and gas masses of this plot 
represent the volumetric values, i.e. they refer to the total star formation rate and 
the total gas mass (i.e. the sum of $M_{HI}$, calculated from Eq. 4,  and $M_{H2}$, calculated from Eq. 5 
of Tab.~\ref{tab_obs}) , and not to the corresponding projected quantities, which are commonly 
observable in local galaxies (see Krumholz et al. 2012). 
As discussed in Sect.~\ref{sect_uncert}, the molecular gas mass could be underestimated if a Galactic 
$L'_{CO(1-0)}$ to $M_{H_2}$ conversion factor were used (Riechers et al. 2006); 
in this case, the observational points would move towards 
the standard model curves. 

For a fair comparison with the projected, local Kennicutt (1998a) law, which 
can be determined for extended (i.e. non point-like) objects, 
one has to divide these SFRs and gas masses by a typical surface. This issue will 
be discussed in more detail later in Sect. ~\ref{sec_sfr}.
 
The discrepancy between the observed $SFR-M_{H}$ relation and the theoretical one 
implies that the systems with observations seem to be 
forming stars more efficiently than our three standard models. 
A similar conclusion can be drawn from the comparison of the $SFR-M_{dyn}$ relation as observed in the 
dataset of Tab.~\ref{tab_obs} and as predicted by our model; for the bulk of the systems with observations, 
at dynamical mass values corresponding to those determined for the QSO hosts, 
our standard models tend to underestimate the SFRs. 

As a final note, as stated in Sect.~\ref{sec_obs}, in this work we are assuming a negligible contribution  
of dark matter to the dynamical mass budget. The presence of dark matter would exacerbate the discrepancy 
between data and theoretical predictions. In quantitative terms, 
for a baryon-to-dark matter ratio of the order of $20\%$ (Fukugita \& Peebles 2004), 
in the $M_{dust}-M_{dyn}$ plot the model curves would 
be shifted towards larger masses by $\sim 0.7$ dex.

\subsection{Effects of star formation efficiency and IMF on dust production} 
\label{res_sfe_imf}
For the systems where observations are available, the observed dust mass and SFR values 
are underestimated by our models presenting dynamical masses and gas masses 
comparable to the observed values. 
This fact 
opens up the possibility for non-standard star forming conditions 
high-$z$ QSO hosts. In this section,  we attempt to quantify the effects of the variation of 
the main SF and dust production parameters on the main observables 
discussed in this paper. \\
In Fig.~\ref{fig_dust_mod}, we show the observed data and the results from three different models, each one 
bearing some changes in the main SF and dust production parameters. 
The magenta dotted lines are the results of a model with a total baryonic mass of 3E11 $M_{\odot}$ and 
with a SF efficiency increased to the value $\nu=60$ Gyr$^{-1}$. 
The cyan solid lines represent a model with both an increased SF efficiency and a decreased 
dust accretion timescale $\tau_{0}=0.002$ Gyr (model M3E11D in Tab.~\ref{tab_models}), which corresponds to larger dust accretion rates, as 
visible in Eq.~\ref{eq_gro}. Finally, the blue dashed lines are for a model with SF efficiency 
and $\tau_{0}$ as above (model M3E11L in Tab.~\ref{tab_models}), but which assumes a modified IMF as suggested by Larson (1998), 
richer in massive stars with respect to the standard Salpeter (1955) IMF. 
In this case, we have assumed a characteristic mass $m_c$ of 1.2 M$_\odot$. \\
An enhancement of the star formation  is required by the 
SFR vs $M_{gas}$ and SFR vs $M_{dyn}$ plots, where the new star formation efficiency helps to improve 
considerably the agreement between model results and data. 
However, the only 
increase of the SF efficiency does not automatically imply that the dust content will be completely 
accounted for; 
in fact, as the $M_{dust}$-$M_{gas}$ and $M_{dust}$-$M_{dyn}$ plots suggest, 
a higher SF efficiency alone causes a low  dust content; this 
happens because of the higher SN rates, which lead also to higher dust destruction rates, as discussed in Sect.~\ref{sec_mod_dust}.

The only way to compensate this effect is by decreasing
the typical grain accretion timescale, boosting the accretion rate: with the value of 
$\tau_{0}=0.002$, the disagreement between models and data is slightly reduced. 

This assumption is a rather extreme one, since it implies a dust growth efficiency an order of magnitude 
higher than the typical value in the solar neighbourhood (Calura et al. 2008). 
A further increase of the dust content is achievable by assuming a top-heavy IMF as suggested by Larson (1998). 
This assumption  
results in an enhanced metal production 
and hence higher rates of production also for stellar dust, as will be discussed in Sect.~\ref{disc}. 
With a Larson (1998) IMF, the dust content of the systems with detections appears reasonably 
accounted for.\\

Fig.~\ref{dtg} shows the dust-to-gas ratio as 
a function of the gas mass and of the dynamical mass as observed in our dataset of QSO hosts and as predicted 
by our models. In the figure, the top panels show the predictions obtained with our 'standard' models M3E10, M1E11 and M1E12, 
whereas the bottom panels of Fig.~\ref{dtg} show instead our predictions obtained with models M3E11S, M3E11D, and M3E11L. 

Most of the observed systems show oversolar dust-to-gas ratios; our models with a Salpeter 
IMF and with standard assumptions for dust tend to underestimate the observed dust-to-gas ratios. 
The assumption of an enhanced star formation efficiency, a more efficient dust condensation and a top-heavy IMF 
helps alleviating the discrepancy between model results and observations in the M$_{dust}$/M$_{gas}$-M$_{gas}$ plot. 
Such assumptions are instead not enough to account for the M$_{dust}$/M$_{gas}$-M$_{dyn}$ relation. 
More extreme assumption for the IMF and higher values for $m_{c}$ do not allow to further 
reduce this tension since in our models, with values for $m_{c}$ higher than 1.2 M$_{\odot}$,  
the increased SN feedback triggers an early outflow and halts the star formation at times earlier than 0.1 Gyr, 
when the dust content is orders of magnitude lower than the values observed in our dataset. 

To summarize, the study presented in this section outlines that, in order to account for 
the dust budget and for the star formation rates observed in high-z QSO hosts, major modifications of a few 
fundamental parameters are required, i.e. an enhanced star formation efficiency, an increased dust growth efficiency 
and a top-heavy IMF. Further implications of these assumptions and possible physical 
motivations are discussed later in Sect.~\ref{disc}.

There is also the possibility that for some systems, the estimated gas mass and the dynamical mass 
may represent a lower limit. This will be the subject of the next section.

\section{Discussion} 
\label{disc}
All the quantities studied in this paper 
are affected by some important uncertainties and systematics effects, which 
need to be taken into account for a fair comparison with model results. \\
On the other hand, from the the theoretical point of view, as seen above in Sect.~\ref{res}, a few main parameters of our models play a fundamental 
role in determining the star formation properties and the dust content of forming spheroids.\\
An enhanced star formation efficiency is required in order to account for the high SFR values 
at low gas masses as observed in high-$z$ QSO hosts. Moreover, 
a high dust growth efficiency is needed to account for the large dust masses observed in such systems. 
Even the stellar IMF plays a fundamental role: an IMF more 
skewed towards massive stars than the local one implies higher rates of metal production, 
consequently a larger rate of CC SNe and of both dust production and destruction. 

In this section, we first discuss a few observational uncertainties affecting the data. 
In the remainder of the section, we discuss instead 
some important implications of our results for the main model parameters analysed in this work. 
We conclude the section by providing predictions on the detectability of the lower redshift 
descendants of the $z=6$ QSO hosts.

\begin{figure*}
\epsfig{file=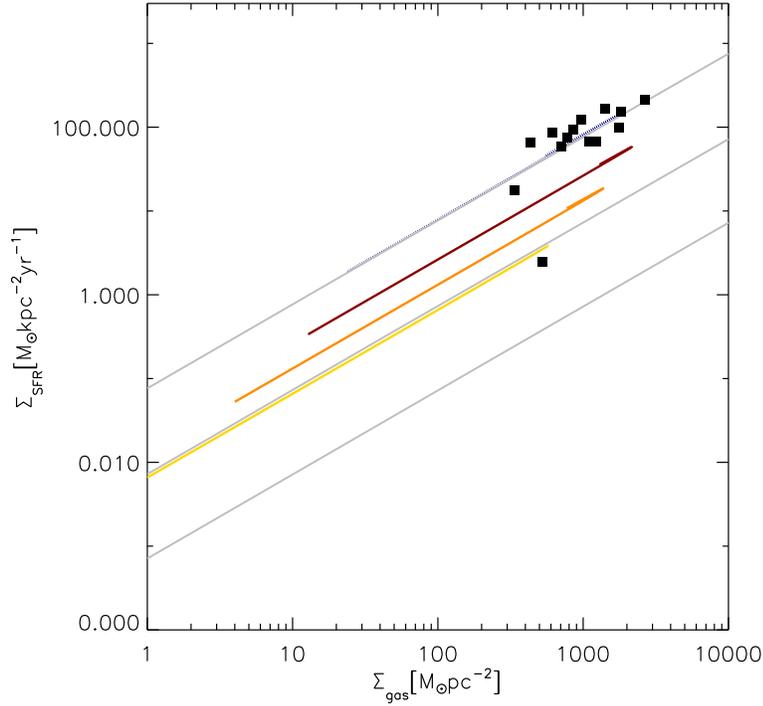,height=10cm,width=10cm}
\caption{Star formation surface density $\Sigma_{SFR}$ as a function of the gas surface density $\Sigma_{gas}$ as observed in high-z QSO hosts 
(solid squares) and as obtained with our models M3E10 (yellow solid line), ME11 (orange solid line), 
 ME12 (dark red solid line), M3E11D (blue dotted line). The thin, grey lines encompass the values derived by Krumholz et al. (2012) 
in local star forming galaxies and in local and high-redshift starbursts. 
To enable the comparison between high-redshift systems and the local relation, 
the star formation rates and gas masses of the QSO hosts are divided by a representative surface value 
$A=\pi D^2$, where $D$ is of the order of a few kpc as 
our fiducial size of the CO emitting region (see Sect.~\ref{sec_obs}).}
\label{sfr}
\end{figure*}

\subsection{Main uncertainties and possible systematics in the data} 
\label{sect_uncert}
Some of the observables studied in this paper are affected by a few major 
sources of uncertainty which need to be pointed out, and which could be  
explained by our models even without invoking non-standard assumptions for the main 
parameters.\\
As discussed in Wang et al. (2013), the dynamical mass is estimated by means of the [C II] and CO line emission, 
assuming that the gas within the emitting region is gravitationally bound and rotating, 
i.e. it lies onto a disc. In these conditions, the dynamical mass can be 
estimated as $M_{dyn}/M_{\odot}\propto v_c^2 D$, where $D$ is the diameter of the disc 
and $v_c$ is the maximum circular velocity. \\
First of all, if some of the systems were not in rotational equilibrium, the observed velocity 
would be rather associable with a velocity dispersion (Tacconi et al. 2006, 2008; 
Riechers et al. 2008; Engel et al. 2010). In this case, the dynamical mass 
would be an overestimate of the real value, owing to non-rotational 
contributions to the detected velocity (Coppin et al. 2010). \\
In addition, some of the systems for which we fixed the disc diameter to 5 kpc could have a smaller size, hence producing
a smaller dynamical mass (see Eq.6). Wang et al. (2013) indeed observed a disc diameter range of 3.3-5 kpc for four $z\sim6$ QSOs observed at 
sub-arcsec resolution with ALMA\footnote{Wang et al. adopted as source diameter D = 1.5 x FWHM of the major axis of [CII] emission}. 
Assuming, as found by Wang et al. (2013), that D$\sim3.5$ kpc in about half of the objects where we
assumed D=5 kpc, this would reduce the $M_{dyn}sin^2i$ estimate by a factor of $\sim1.4$ in only three objects out of 13 for which $M_{dyn}sin^2i$
is available. This small systematic uncertainty is not likely to bias our results significantly.

However, there are also several reasons for which the real 
dynamical masses could have been underestimated.  

First of all, in principle, the detected emission could allow one to trace only the brightest, innermost regions 
of the QSO hosts. If, for some systems, a non-negligible amount of the gas were distributed around this central region 
and characterized by a low surface brightness, its detection would be problematic and 
the dynamical mass could be severely underestimated. 
For this reason, the different angular scales over which the gas and dust content are
measured is a source of concern. 
Indeed, most FIR (dust) measurements come from MAMBO at an angular resolution of 11 arcsec
($\sim60$ kpc at z=6), while PdB and ALMA observations of CII and CO lines have even sub-arcsec resolution.
All measurements by MAMBO, however, are unresolved, and, in the few cases where both MAMBO and ALMA
continuum observations were available together with ALMA CII emission (Wang et al. 2013), 
it was shown that all the dust emission measured by MAMBO on scales $<60$kpc 
is in reality co-spatial with the gas within a few kpc region.

However, recent CO observations found no evidence 
for extended reservoirs of low-surface brightness 
molecular gas beyond the compact nuclear region of high-$z$ QSO host galaxies 
(Riechers et al. 2006, 2011). On the other hand, a low surface brightness component 
is found in some submillimetre-selected galaxies at $z=2.2-2.5$ (Ivison et al. 2011), 
outlining  the possible existence of large non-star-forming molecular gas reservoirs beyond the star-forming region. 

In addition, the conversion factor $\alpha$ used to derive molecular gas masses from 
measured CO emission is generally valid in star-forming galaxies similar to 
local ULIRGS/starbursts (Downes \& Salomon 1998). 
If a faint extended component were present, i.e. if a value for $\alpha$ more appropriate for Milky-Way like galaxies 
were used, the estimated $H_{2}$ mass would be larger. 

As far as our systems are concerned, there is a possibility that
the CO emission could be more extended than the [C II] emission. This cannot be
ruled out, e.g., for J1148+5251, based on existing observations
(Walter et al. 2009; Riechers et al. 2009).

If this were the case of our systems, i.e., by taking into account 
a substantial contribution of undetected gas in the gas mass estimate, 
the observed SFR-$M_{gas}$ relation would move towards the region occupied by our standard models, 
without requiring  a SF enhancement as for model M3E11S. 
However, the contribution of this non-star forming gas to the dynamical mass 
would leave less room for the stellar masses as calculated by means of Eq.~\ref{eq_mdyn}, 
hence the dust yield per unit stellar mass required to explain the large dust masses in the $M_{dust}$-$M_{star}$ plot
would be even larger, exhacerbating the discrepancy between data and models.

Another major source of uncertainty is the inclination angle of the disc component. 
If a rotating disc is present in the observed system, 
the true observable to determine the dynamical mass is $v_c \propto 1/sin\,i$ (Wang et al. 2010), where 
$i$ is the inclination angle of the disc with respect to the line of sight. 
As reported in Tab.~\ref{tab_obs}, the 
directly derived quantity is then $M_{dyn} sin\,i^2$; the estimated inclination 
angles indicate values $24^{o}\le i \le 56^{o}$, implying modest corrections to the 
dynamical masses (Wang et al. 2013). 
For comparison, for randomly inclined galaxy discs one would expect an average 
inclination $i=30^o$ (Coppin et al. 2010).  \\
In a few systems, the inclination angle has been recently estimated from the [C II] minor and major axis ratios, 
which still have large uncertainties (Wang et al. 2013). 
However, in other systems, the possibility of small inclination angles cannot be completely ruled out, 
and this could cause the dynamical mass to be considerably underestimated. 

Each of these sources of uncertainty in the observationally derived gas masses and dynamical masses could 
be a possible reason for the discrepancy between the results of our standard models and the data. 
If for some systems 
the atomic gas masses or the molecular gas masses represent underestimates of the total amount of 
star-forming gas, 
their SFR values and dust masses should not be compared with our intermediate mass model 1E11, but with a more massive one. 
Interestingly, for the systems characterised by detections, 
the dust masses and SFR values of our most massive model 1E12 are compatible with the observed ranges; 
however, 
this model would be a faithful description of the high-$z$ QSO hosts only if, for most of them, the 
gas and dynamical masses had been underestimated by a factor of $\sim 10$. 
It is also worth to stress that in this case,  
no modification of the IMF, SF efficiency or dust growth timescale would be required. 
Even if it is unlikely that such a large underestimation of the masses 
may affect all the systems of Tab.~\ref{tab_obs}, the possibility that it could concern  a non-negligible 
fraction or even the majority of the dataset cannot be completely excluded. 

Moreover, we note that most measurements of $L_{FIR}$, SFR and $M_d$ in our sample are based on a single datapoint
at 250 GHz from MAMBO, and therefore heavily depend on the assumed FIR spectral energy distribution (SED).
As discussed by Omont et al. (2013), for a given flux density at 250 GHz, the values of $L_{FIR}$ (and SFR)
would be $\sim3$ times lower if $T_d=33$ K is assumed instead of $T_d=47$ K. On the contrary,
$M_{dust}$ would be $\sim2.4$ times larger. \\
Other systematics could affect the estimate of the dust masses: 
incorrect calibrations for conversion factors between IR luminosities and dust masses could lead mainly to 
overestimate the dust content of high-$z$ starbursts (Mattsson 2011), 
possibly owing to an higher average dust emissivity (Santini et al. 2010). \\
Also the SFR values could be overestimated, mostly owing to dust grain heating by 
sources different from massive stars, such as 
an active galactic nucleus (AGN, see Omont et al. 2013); 
no significant contribution is however expected at 250 GHz. 
At present, measures of the AGN contribution to the dust continuum 
in high-$z$ systems are being planned with ALMA (Wang et al. 2013, Gilli et al. 2013).\\
In principle, also intermediate-age stellar populations could contribute to the FIR emission, 
as occurs in IR luminous galaxies at $z<2$ (Lo Faro et al. 2013).  
However, given the likely young age of the systems studied in this paper, we 
do not think that their presence can provide an important contribution to the mid-far infrared flux. 
Moreover, the denser environments and the different physical conditions of high-z starbursts 
could render the local calibrators unsuited to derive the SFR at high redshift (Fan et al. 2013). \\

\subsection{The star formation efficiency} 
\label{sec_sfr}
The systems studied in our paper are forming stars with particularly 
high efficiencies, even higher 
than what implied by the basic scaling relations of the most massive local ellipticals in which, 
to account for the observed integrated [$\alpha$/Fe]- stellar mass relation, 
SF efficiencies of the order of 
$\sim 20$ Gyr$^{-1}$ are required (Pipino et al. 2011 and references therein). 
The implied star formation timescales (see Fig.~\ref{fig_models}) are less than half the values 
commonly believed as typical for spheroids of comparable final stellar mass ($\sim 10^{11} M_{\odot}$) 
i.e. approximately 1 Gyr (see Renzini 2006; Lapi et al. 2011 and references therein).  

The self-regulating nature of the star formation process 
can be appreciated even in extreme conditions as in these systems: 
despite their very large SF surface densities (Walter et al. 2009), 
stars cannot form much more rapidly than they currently are without
removing the remaining gas due to radiation pressure, hence without 
quenching the SF activity. 

In this section, we aim at assessing whether the systems studied in this paper 
can still be considered as part of a universal star formation 
sequence implied by the Kennicutt-Schmidt law as derived in local and distant discs and starbursts (Krumholtz et al. 2012), 
which seems to
hold up to high redshift and be valid for systems
presenting a  large range of densities and star formation rate values
(Bouch\'e et al. 2007; Krumholtz et al. 2012). \\
By dividing the 
star formation rates and gas masses by a representative surface value 
$A=\pi D^2$, where $D$ is of the order of a few kpc as 
our fiducial size of the CO emitting region 
(see Sect.~\ref{sec_obs}),  we can switch from a
'volumetric' star formation law to a 'projected' one,  which is more
easily comparable with the Kennicutt (1998a) law observed in local
star-forming galaxies. Our projected model SF rates as a function of
the gas surface densities are reported in Fig.~\ref{sfr},  together with a
local SF law encompassing a large variety of star forming systems at
low and high redshift (Krumholtz et al. 2012).  \\
As far as the observational sample of Tab.~\ref{tab_obs}, 
here we limit our analysis only to the systems with detections for both the gas mass and the SFRs.

Fig.~\ref{sfr} shows that the SF efficiencies assumed for models M3E10, M1E11, M1E12 
produce SF rates which agree with a local, projected star formation law. \\ 
On the other hand, most of the observational data and 
our model M3E11S lie on the extreme upper edge of the universal star formation law as derived by 
Krumholz et al. (2012). This may be regarded as the result of a selection 
effect, since our sources are selected mostly on the basis of their high FIR luminosity. 
Notwithstanding this aspect, Fig.~\ref{sfr} shows also that most of the observational data lie 
within the scatter of the universal Kennicutt-Schmidt SF law as derived by Krumholz et al. (2012). 

A very high star formation efficiency
in large spheroids, such as the one  assumed here for our model M3E11S,
is explainable in galaxy evolution models taking  into account
a positive (i.e. boosting SF) feedback produced in the early stages
of supermassive central black hole  growth (Silk 2005; Ciotti \& Ostriker 2007; Pipino, Silk \& Matteucci
2009, Silk \& Norman 2009).  
In such a model, SF is triggered coherently and rapidly by a
SMBH jet-induced hot plasma cocoon, which sweeps up and over pressures cold clouds (Silk 2005). 
The instabilities caused by a 
broad jet propagating into an inhomogeneous ISM leads cold clouds to shock 
and disperse into filaments (Krause \& Alexander 2007; Antonuccio-Delogu \& Silk 2008). 
Over time, more and more gas condenses into the cold phase and fragment 
into small stable cloudlets, which may host SF (Mellema et al. 2002; Fragile et al. 2004). 
The overall feedback to SF is positive, and 
the interaction of the
outflow with the surrounding protogalactic gas at first stimulates SF
on a time-scale which is much shorter than the one required by negative feedback from SNe 
to take place. Such a scenario is supported by evidence 
for jet-stimulated SF in high-z systems (Bicknell et al. 2000; Venemans et al. 2004), followed
in at least one case by a starburst-driven superwind (Zirm et
al. 2005). At low-redshift, an example of jet-induced starburst is  
Minkowski's object (van Breugel et al. 1985; Croft et al. 2006). 

One might ask how could a scenario where star formation is enhanced 
by the presence of an AGN be reconciled with the existence of galaxies observed at $z>6$ showing similar SFRs 
and without evidence for a luminous AGN (e.g., Riechers et al. 2013). 
According to recent studies based on numerical simulations and taking into account 
AGN-induced star formation, the effects of the SF enhancement may last for timescales 
up to 10 times longer than the duration of the AGN activity ($10^6$-$10^7$ yr, Zubovas et al. 2013).
Because of the short duration of the AGN 
phase, the jet-induced SF-enhancement  may be more common than 
the detections of systems presenting simulaneously a starburst and 
an AGN. 

The enhancement of the SF could also be caused by merging activity, in this case without 
requiring a positive AGN feedback (Valiante et al. 2011; 2012; Khandai et al. 2012). 
In this scenario, one major difficulty is to explain the undisturbed morphologies 
as visible in a large fraction of the systems observed at $z\sim 6$ (Wang et al. 2013): 
the last major merger triggering the starburst must date back enough 
to allow the systems to relax and present an undisturbed velocity profile. 
Another major difficulty of explaining the properties of the most massive QSO hosts at $z\sim6$ by means of 
mergers, in particular within LCDM galaxy formation, is that LCDM-based models 
generally tend to underpredict the rate of merging, hence the abundance, 
of massive galaxies at high redshift (Conselice 2006). 
Moreover, within the same scenario, various results have shown that a strictly negative feedback of the 
AGN is not enough to accelerate the gas consumption timescales enough to account for the 
basic scaling relations of elliptical galaxies outlining their marked anti-hierarchical mass buildup, 
such as the [$\alpha$/Fe]-mass relation (Pipino et al. 2009b; 
Calura \& Menci 2011; Yates et al. 2013). 
It seems instead that a positive effect of the AGN feedback, quickening the star formation timescales 
in the most massive systems, is fully compatible also with the 'downsizing' character shown both by distant 
galaxies (Cowie et al. 1996; Kodama et al. 2004) and by the archeological record of the stellar populations of 
local early-type galaxies (Matteucci 1994). 
At present, growing evidence is emerging that secular processes may naturally lead to
significant AGN and starburst activity, even in absence of external phenomena such
as galaxy merging (Genzel et al. 2008; Ciotti \& Ostriker 2012 and references therein).

\subsection{Implications for dust production at high redshift}
\label{sec_dust} 
Large dust masses characterise  not only the QSO hosts  
studied in this paper, but also dusty starbursts at $z>6$, which
may be as common, and may contain even more dust (e.g., Riechers et
al. 2013). To explain the large dust content of high-z starbursts, 
rather extreme model assumptions are required. 
As discussed above in Sect.~\ref{sec_sfr}, a 
high star formation efficiency implies large SN rates, hence large dust destruction rates. 
In order to enhance the dust production and 
account also for the dust masses at low gas and dynamical masses, 
a viable solution is to decrease the dust accretion parameter $\tau_{0}$, 
i.e. increasing the dust growth efficiency. \\
In principle, this quantity should be a complex function of 
various parameters related to the micro-physics of dust grains, such as 
the grain size, the local cloud temperature and density, and the interstellar metallicity 
(see Dwek 1998; Zhukovska et al. 2008; Pipino et al. 2011; Kuo \& Hirashita 2012).  
A single-zone, single-phase model such as ours does not allow us to have 
a deep insight into the true dependence of $\tau_{0}$ on each of these parameter. 
However, our models are useful to roughly constrain this parameter 
and to test whether a value suited for the local dust budget is 
acceptable also for high-z systems, likely characterized by significantly  
different physical conditions.\\
Our models show that, if very high star formation efficiencies need to be assumed for high-z QSO hosts, a 
value for $\tau_{0}$ of the order of $\sim 0.01$ Gyr, i.e. a factor of  2 shorter than the local 
typical molecular cloud lifetime, is not sufficient to account for their  dust  content. 
Acceptable results are instead achieved assuming 
$\tau_{0}=0.001$ Gyr, i.e. an order of magnitude smaller than the one assumed in chemical evolution models 
designed to reproduce the properties of local galaxies. This assumption is 
supported by other theoretical attempts to model the dust content of galaxies. 
Sophisticated models including a treatment of the grain size distribution 
and physical processes such as shattering (Kuo \& Hirashita 2012) and coagulation (Asano et al. 2013) 
indicate that the grain size has an important effect on the grain growth. 
In particular, at extreme star forming conditions such as in high-z starbursts, 
turbulence-driven collisions between grains moving at high velocities could render 
grain shattering very efficient, with the production of large numbers of small grains.

The grain growth rate is an increasing function of the surface-to-volume ratio of 
the dust grains: an efficient shattering increases this ratio, enhancing the accretion process (Kuo \& Hirashita 2012). 
According to these results, a  
shorter accretion timescale in starbursts could then be explained by a particularly large abundance of 
small grains. However, this is in contrast with the flat extinction curves derived from 
optical-near infrared spectra of high-$z$ QSOs (Gallerani et al. 2010), implying an overabundance 
of large grains. However, in this case an enhancement of dust growth could be 
due to larger cloud densities. In fact, as  
mentioned in Kuo \& Hirashita (2012), also larger average particle densities in high-z molecular 
clouds are likely to decrease the accretion timescales, as well as the supersolar metallicities 
which may characterise some QSO hosts (D'Odorico et al. 2004). \\
Various groups attempted to study the dust content of high-z QSO hosts by means of theoretical models, 
usually finding that several sources could be responsible for their large dust content. 
Dwek \& Cherchneff (2011) point out that also AGB stars may have given an important contribution 
to the dust mass of QSO hosts at $z\sim 6$, but only after ages of $\sim0.4$ Gyr, 
showing that ad-hoc star formation histories can be conceived where even type II SNe could be regarded as 
important dust producers. 
However, the dust yields per unit stellar mass required by the large dust content of $z\sim 6$ QSOs 
are probably too high even for a top-heavy IMF (Michalowski et al. 2010). 
Most of the previous attempts to reproduce the dust masses of high-$z$  
starbursts support a basic need for an alternative mechanism to stellar dust production 
(Pipino et al. 2011; Valiante et al. 2011; Mattsson 2011), 
unless a non-standard stellar mass function is assumed (Gall et al. 2011).

\subsection{The role of the stellar initial mass function} 
\label{sect_imf}
Another fundamental parameter in chemical evolution models is 
the stellar initial mass function, which could be different from the local one in 
environments characterized by peculiar physical conditions, such as high-$z$ starbursts. 
Larson (1998) presented a list of circumstantial evidence supporting the possibility 
of a high-$z$ IMF more skewed towards massive stars with respect to the local one. 
First of all, the cosmic microwave background temperature is a linear function of redshift; consequently, its 
value at redshift $z\sim 6$ could be higher with respect to local by  
up to a factor of $\sim 7$. This temperature sets 
the minimum molecular cloud temperature; a larger value for this quantity 
implies a larger Jeans mass, hence larger stellar masses if the latter 
can be regarded as a characteristic mass in the star formation process (Narayanan \& Dav\'e 2012). 
Furthermore, larger average molecular cloud masses can also be explained by 
a less efficient cooling at lower metallicities and by means of a particularly intense UV radiation field. 
Also the 'G-dwarf' problem, i.e. the overabundance of low-metallicity stars 
predicted by closed-box chemical evolution models using a standard IMF, 
may be partially alleviated with the assumption of an IMF poor in low mass stars, even though 
other realistic solutions could be proposed, such as an infall of pristine gas. 
Other indirect evidence involve the observed 
evolution of the optical luminosity density, which may be underestimated using a normal IMF if 
dust extinction is taken into account. Also the far-infrared background radiation 
seems to be explainable by means of a bimodal star formation with an enhanced formation of massive stars 
in high-$z$ starbursts (Dwek et al. 1998), as well as the discrepancy between 
the observed present-day stellar mass density and the integral of the comoving star formation rate 
density (Wilkins et al. 2008; Dav\'e 2008).  

\begin{figure*}
\epsfig{file=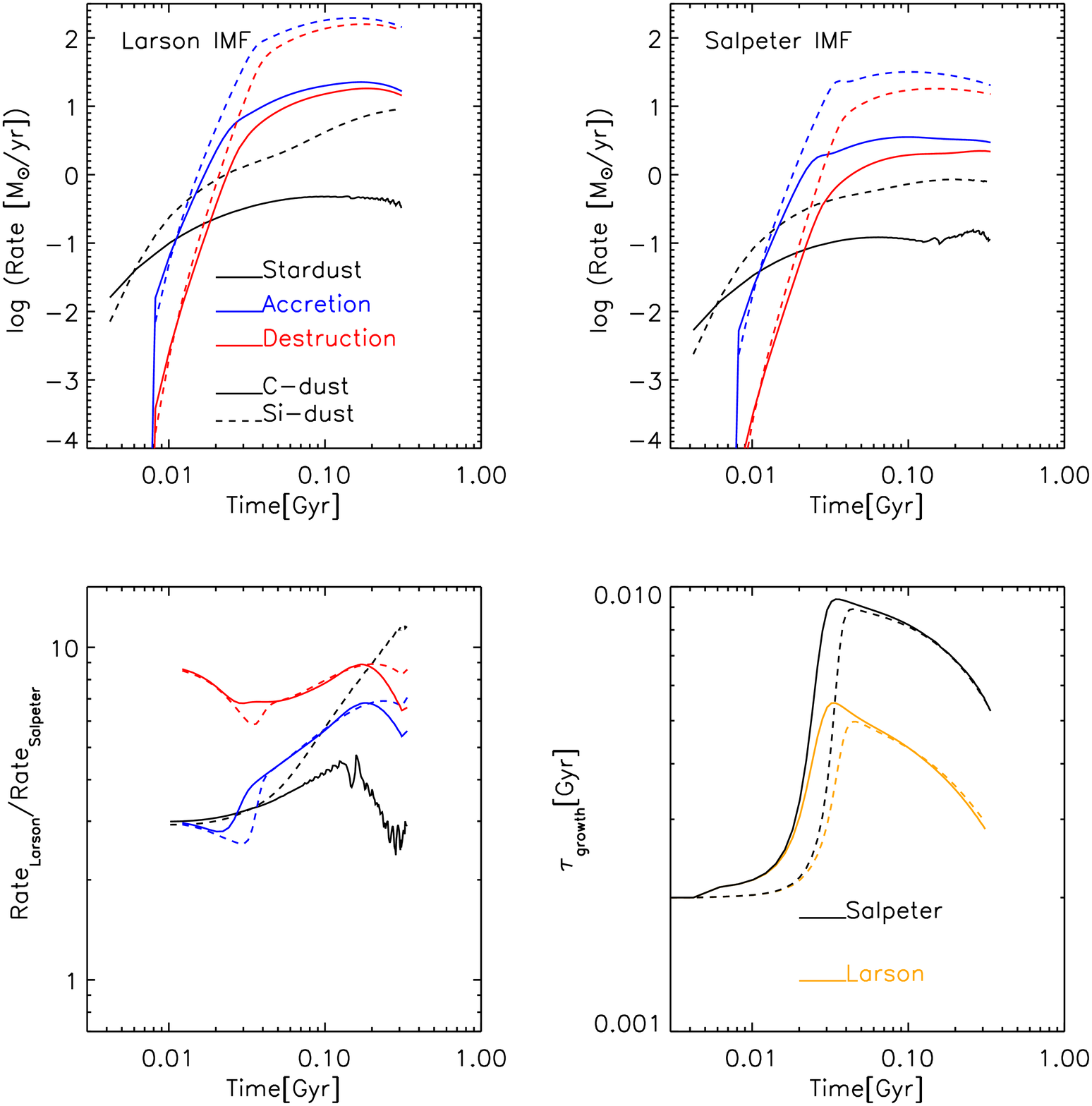,height=18cm,width=18cm}
\caption{Top panels: production and destruction rates for 
carbon (solid lines) and silicate (dashed lines) dust as 
a function of time.  
The red lines represent the destruction rates, the blue lines the accretion rates and the black lines 
represent the total stardust production, i. e. including AGBs and SNe. 
In the top-left panel, a Larson (1998) IMF is assumed, with a characteristic mass $m_c$=1.2 M$_{\odot}$. 
In the top-right panel, a Salpeter (1955) IMF is assumed. The bottom-left panel shows the time evolution of the ratios between production and destruction rates as obtained using a Larson and a Salpeter IMF (lines 
and colours as in the top-panels). The bottom-right panel shows the time evolution of the dust growth timescales 
calculated for C dust (solid lines) and for Si dust (dashed lines) assuming a Larson (orange lines) and 
a Salpeter IMF (black lines). } 
\label{fig_rates}
\end{figure*}

Although no direct, ultimate proof for a different IMF in the early 
Universe can be found, a top-heavy IMF offers a tantalizing explanation 
also for the large dust masses observed in high-$z$ starbursts (Gall et al. 2011). 
In this paper, we have considered the effects of a Larson IMF with 
a characteristic mass $m_c=1.2$ M$_{\odot}$, 
larger than the local Jeans mass ($M_{J}\sim 0.3$ M$_{\odot}$, Bate \& Bonnell 2005). 
The IMF assumed for model M3E11L produces a  
sufficient enhancement in dust production to account for the dust masses 
observed in high-$z$ QSO hosts. The reason may be understood by looking at Fig.~\ref{fig_rates}, 
where we show the dust production and destruction rates calculated assuming both a 
Salpeter and a Larson IMF. 
The use of a Larson IMF yields an increase of the production of both carbon and silicates grains 
with respect to a Salpeter, accompanied by an increase of the destruction rates, owing to 
considerably larger core-collapse SN rates. 
However, in our model, with a Larson IMF also the dust growth rate increases, as 
visible by the comparison between the growth time scales achieved in the two cases 
(Fig.\ref{fig_rates}, bottom-right panel). \\
The larger dust accretion rates obtained with a Larson IMF 
may be explained with lower fractions of refractory elements into dust, due primarily to more efficient 
dust destruction by SNe. Even if the choice of a Larson IMF enhances the metal and dust production, 
the dust-to-metals ratio decreases, and this causes shorter dust accretion timescales. \\
Dwek et al. (2011) have stressed the non-uniqueness of a solution 
invoking a top-heavy IMF for reproducing the dust content of high-$z$ starbursts, outlining 
the role of other mechanisms, including the relative balance between the dust destruction and 
the dust growth efficiencies, in determining the dust content of high-redshift galaxies. The same authors 
have outlined how a reliable determination of the stellar mass may help limiting the space of the parameters, 
in particular the role of the IMF. An estimate of the stellar mass via 
SED fitting is feasible for high-$z$ starbursts around obscured QSOs (Gilli et al. 2013).\\
As seen in Fig~\ref{fig_models}, a top-heavy IMF as the one assumed here has a deep impact also on the interstellar metallicity. 
In principle, the detection of emission lines 
of interest for the determination of the metallicity, such as [OIII], H$\beta$, [OII] could help achieving 
fundamental constraints on the IMF in high-redshift systems. \\

\subsection{Implications for the SMBH-stellar mass ratio} 

As discussed in Wang et al. (2013), there are currently 
more than  60 systems at $z\sim 6$ for which it has been possible to derive 
reliable estimates of the mass of the central supermassive black hole (SMBH). 
In general, SMBH masses range from $10^8$ up to $10^{10}$ M$_\odot$
(Kurk et al. 2007; Jiang et al. 2007; Willott et al. 2010a; De Rosa et al.
2011), suggesting efficient black hole accretion within 1 Gyr after the Big Bang (Di Matteo et al. 2012). 

The dynamical masses estimated from CO in 
high-$z$ starbursts imply 
mass ratios between the central SMBHs  and the spheroidal bulge
an order of magnitude higher than the mean value found in local spheroidal
galaxies (Walter et al. 2004; Reichers et al. 2009), which is typically $2\times 10^{-3}$ (Marconi \& Hunt 2003). 
As outlined in Wang et al. (2013), this evidence was explained as 
an early phase of galaxy evolution, in
which the SMBHs are accreting at their Eddington limit, whereas the quasar
stellar bulges are still accumulating their mass via massive star formation.

As stressed by Wang et al. (2013), 
the detected [C  II] line emission may trace only the intense star forming region 
in the very centers of quasar host galaxies and not extend as far as the stellar bulge. 
However, the stellar bulges of QSOs at $z>6$ are unlikely to be
significantly more extended than the measured sizes of the [C II]
emission, since the hosts are not detected beyond the point-spread function in deep HST imaging. 
Additionally, the large error bars of the deconvolved
minor axis and position angle measurements of some of the objects bear significant 
uncertainties in the inclination angle and in the dynamical mass estimates. 

Our results show that, to account for 
the dust masses observed in high-$z$ starbursts, with standard assumptions regarding 
the model parameters, larger stellar masses than those inferred from observational data  
are required. 
On the other hand, our results have also shown that 
the large dust-to-stellar mass values can be explained with a 
top-heavy IMF, which could represent an explanation also for the large $M_{BH}/M_{bulge}$ 
estimated for high-$z$ QSO hosts.

\subsection{On the detectability of the descendants of high-z QSO hosts} 
The intense, dust-enshrouded  starbursts identifiable with our systems  
are likely to represent the progenitors of massive galaxies (i.e. with $M_{*} \simgt 10^{11} M_{\odot}$). 
This is also supported by our results;  
the production of large amounts of dust has necessarily to be accompanied by 
large stellar masses. On the other hand, such starbursts cannot last for periods much longer than 
$\sim$ 0.5 Gyr. After such period, sufficient to produce masses of the order of $\sim 10^{11} M_{\odot}$, 
the intense starbursts originate  massive outflows, 
halting the star formation. 
In principle, this means that the descendants of some of 
our systems may be potentially visible in the form of large, passive systems. 
It may be of interest to assess whether the sensitivity limits of future surveys may allow us to detect 
such large, passive systems at high redshift. 

By means of a spectrophotomoteric code suited to generate spectra of composite stellar populations (Jimenez et al. 2004), 
starting from the star formation history of model M3E11D, 
we have calculated the evolution of the spectral energy distribution of the model galaxy. 
The flux emitted at the wavelength $\lambda$ at the time $t$ by a composite stellar population  is:
\begin{equation}
F(\lambda,t)=\int_0^t\int_{Z_0}^{Z_f}\psi(t-t') F_{SSP}(\lambda,Z,t')dt'dZ, 
\end{equation}
where $F_{SSP}(\lambda,Z,t')$ is the flux of a simple stellar population, characterised by a 
metallicity $Z$. The integral above is evaluated between $Z_i$ and $Z_f$, representing the initial and final 
metallicity of the star-forming gas, respectively. 

\begin{figure*}
\epsfig{file=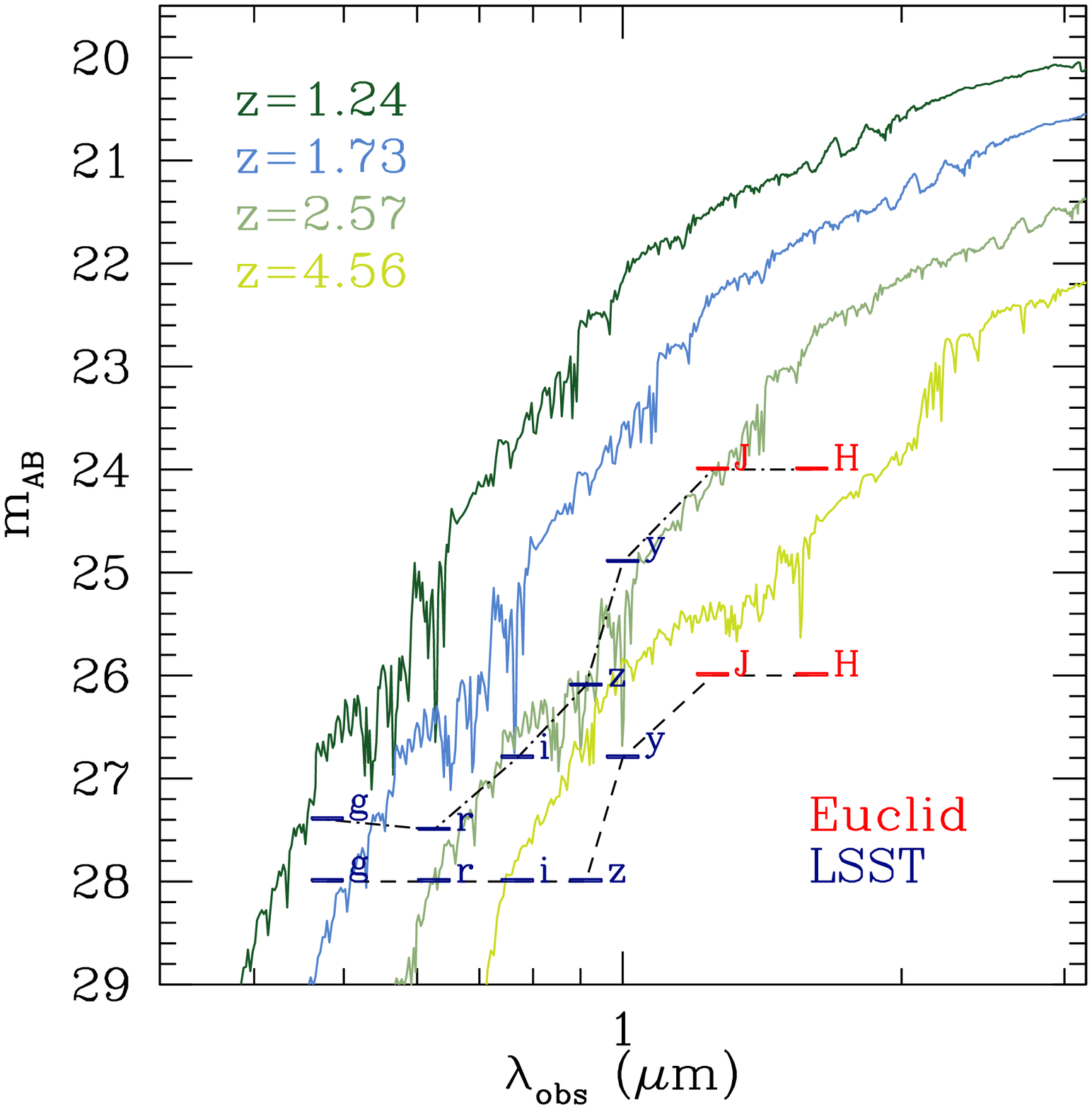,height=18cm,width=18cm}
\caption{Expected optical/near-IR spectra of the low-redshift descendants of a system hosting a $z\sim6$ QSO.
A dynamical mass $M_{dyn}=3\times10^{11} M_{\odot}$ and a Salpeter IMF have been assumed here (model M3E11S in Table 2).
At $z<5$ both star formation and QSO activity have ceased, and the spectra are dominated
by passively evolving stars. The sensitivity limits of future surveys performed by LSST and Euclid are shown
as horizontal ticks. Upper (lower) ticks connected by a dash-dotted (dashed) line show the magnitude limits to be reached by the
LSST and Euclid wide (deep) surveys. Passive systems can be detected and identified as early as $z>4$ in the deep portion of those surveys. } 
\label{fig_sed}
\end{figure*}
 
In Fig.~\ref{fig_sed}, the SEDs computed at various redshifts for our model M3E11D are plotted, 
together with the magnitude limits of two reference future instruments, 
the Large Synoptic Survey Telescope
(LSST Science Collaboration et al. 2009)  and Euclid (Laureijs 2011). To compute the SEDs in Fig.~\ref{fig_sed}, for 
our model galaxy we have assumed a 
formation redshift of $z_{f}=10$. 

These assumptions imply that the systems observed at $z\sim 6$ should be 
$\sim 0.5$ Gyr old; according to the star formation history of our model ME311S, at 0.5 Gyr 
star formation is about to cease. 
Fig.~\ref{fig_sed} shows that, in principle, the massive descendants 
of the QSO hosts could be detected by both LSST and Euclid out to al least $z\sim 2.5$. 
They could even 
be detected at larger redshifts ($z\sim 4.56$) thanks to the deep portions of these surveys.

In order to assess how many of such systems will be observable by future surveys, 
we can evaluate the expected integral number  
$N(<m)$ of descendants of QSO hosts brighter than $m$. 

In a given redshift interval between $z_{min}$ and $z_{max}$, 
the quantity $N(<m)$ can be calculated from the integral
\begin{equation}
N(<m)=\int_{z_{min}}^{z_{max}} \int_{M_{min}}^{M_{max}(m,z)} \Phi(M,z) \, dM \, \frac{dV}{dz}dz,  
\end{equation}
where $\Phi(M)$ is the luminosity function (LF) and  $dV$ is the comoving volume element (Pozzetti et al. 1996; Shimasaku \& Fukugita 1998). 
The relation between the absolute magnitude $M$ and the apparent magnitude $m$ is 
\begin{equation}
M(m,z)=m- 5\, log\frac{d_L}{10} + E(z) + K(z), 
\end{equation}
where $d_L$ is the luminosity distance expressed in pc, $E(z)$ is the evolutionary correction and $K(z)$ is the K-correction.

\begin{figure*}
\epsfig{file=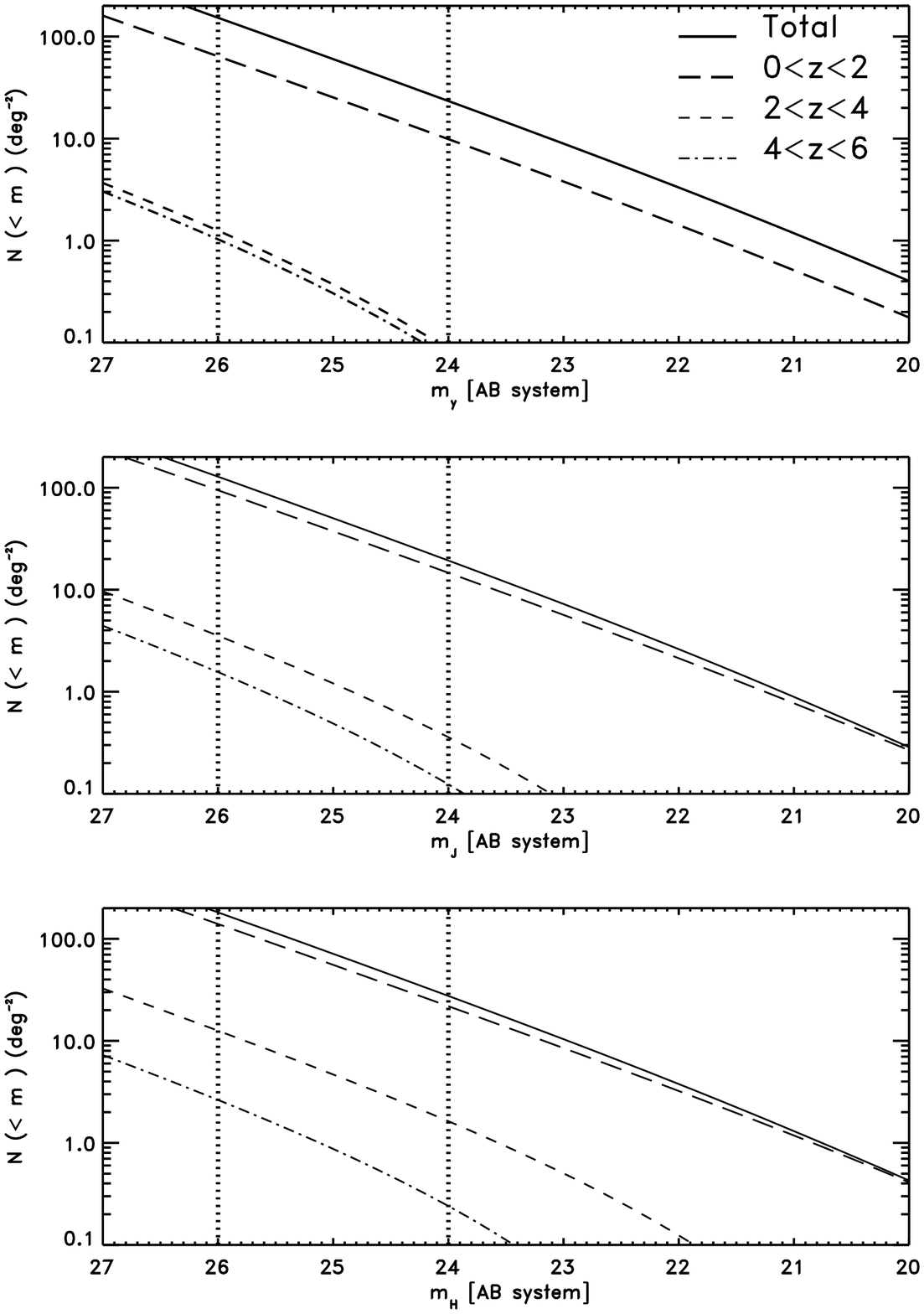,height=18cm,width=18cm}
\caption{Predicted integrated counts of massive QSO descendants in various redshift bins and in three Euclid bands 
(from top to bottom: Y, J and H band). The counts have been generated by means of a SED calculated for the M3E11S 
model and the BH mass function determined at $z=6$ by Willott et al. (2010b). 
In each panel, the solid line represents the total integrated counts, whereas the long-dashed line, 
the short-dashed line and the dot-dashed line are the counts at redshifts $0<z<2$, $2<z<4$ and $4<z<6$, respectively. 
The dotted vertical lines at $m=26$ and $m=24$ correspond to the deep and wide Euclid magnitude limits, respectively. } 
\label{fig_counts}
\end{figure*}

To assess the luminosity function of the massive QSO descendants, we use the 
black hole mass function estimated by Willott et al. (2010b) 
for a sample of quasars at $z\sim 6 $. 
This function is converted into a LF by assuming a fixed $M_{BH}/M_{bulge}=0.0014$, i.e. 
corresponding to the local value (Marconi \& Hunt 2003), by computing the $M_{bulge}/L$ ratio directly 
from our SEDs, and assuming that the latter quantity is does not vary with the 
bulge stellar mass $M_{bulge}$. 

In Fig.~\ref{fig_counts}  we show the expected integrated counts in various redshift bins and 
in the three Euclid near-infrared bands $Y$, $J$ and $H$. The dotted vertical lines in Fig.~\ref{fig_counts} 
correspond to the magnitude limit of the deep and wide Euclid surveys, i.e. AB magnitudes 26 and 24, respectively. 

The largest contribution to the number counts come from the systems at redshift between 0 and 2. 
In the H band, $\sim 182$ systems per square degree are expected at $m_H=26$; this means that in the area of 
the Euclid deep survey  of 40 square degrees, $7280$ objects are expected. 
The Euclid wide survey has a considerably larger area, of the order of 15000 $deg^2$ 
(Laureijs 2011; Majerotto et al. 2012); in such an area, $\sim 3.8 \, 10^5$ QSO host descendants 
are expected between $z=0$ and $z=2$. 
More problematic is the detection of such objects at larger redshifts: 
in the J and H bands, $\sim 60$ and $\sim 120$ objects between $z=4$ and $z=6$ are expected in the deep portion 
of the Euclid survey, respectively.  

Figures ~\ref{fig_sed} and ~\ref{fig_counts} show that the sensitivity of future wide-field surveys is sufficient to provide a complete characterization of the optical to near-IR SED of the descendants of  $z\sim6$ QSO hosts.
Therefore, although their surface density on the sky will be overwhelmed by that of other galaxies and local stars,
accurate SED-fitting procedures would provide reliable photo-z measurements to identify these descendants. 

It may be interesting to compare our 
predictions to actual depths and areas reached in existing surveys, i.e. COSMOS and CANDELS. 
Overall, given their small areas, we do not expect to detect a significant number of
descendants of  $z\sim 6$ QSOs. 
In the COSMOS UltraVISTA survey (McCracken et al. 2012, with an area of $\sim$1.5 $deg^2$), 
at H-band magnitude 25 
we expect to detect $\sim$100 QSO
descendants in total, with only 1-2 of them at very high-z ($4\le z \le 6$). 
Similar numbers are expected in the deep pencil-beam CANDELS survey (0.2  $deg^2$, Grogin et al. 2011) 
in the H band at magnitude 27. 

Finally, it is worth noting that if we had assumed a Larson (1998) IMF to compute the spectra, 
the detectability of the 
high redshift descendants would be more problematic; in fact, such an IMF causes lower stellar masses and consequently 
fainter objects. \\
Also the assumption of a larger $M_{BH}/M_{bulge}$ ratio as suggested by some authors (Wang et al. 2013)   
would worsen their detectability, since it would imply a shift of the mass function towards 
fainter magnitudes, i.e. a lower number density at fixed stellar mass.

\section{Conclusions} 
The dust content of high-$z$ QSOs is a record of 
their star formation history. 
In this paper we have shown how, by 
studying the dust content of such systems in combination with other observables, 
such as the neutral gas mass, the molecular gas and the SFR, it is possible to achieve  
crucial information on their star formation efficiency and derive constraints for other 
fundamental parameters, such as the dust accretion timescale and their  initial stellar mass function.\\
In this paper, we have collected a database of 58 high redshift ($z \gtrsim 5.7$) QSOs from various sources in the literature and  
discovered by various surveys, including the SDSS, the CFHQS, the NDFWS and the UKIDSS survey and observed 
in the FIR. 
For each systems, by means of standard calibration methods, we derived the dust masses, the SFRs, 
and, when possible, the neutral gas masses and the H2 masses. 
In a few cases, estimates of the dynamical masses are available.\\ 
We have interpreted the observed dust masses, gas masses and dynamical masses by means 
of chemical evolution models designed to reproduce the main features of local elliptical galaxies. 
These models are tuned to account for the local abundance ratios (e.g., Pipino et al. 2009a) and the 
dust content of local early-type galaxies (Calura et al. 2008), and they have 
already been used to study the dust budget in one QSO host galaxy at $z=6.4$ 
(Pipino et al. 2011). We have considered chemical evolution models for galaxies of three different 
baryonic masses, encompassing the range of dynamical masses derived for high-$z$ QSO hosts. 
Some fundamental parameters regulating star formation and 
dust production, including the initial stellar mass function, 
have been studied in detail when comparing model results with observed data. 
Our conclusions can be summarized as follows:\\
\begin{itemize}
\item Most of the observed systems present upper limits for the dust masses. For such systems, 
in general the dust masses, gas masses and SFRs are consistent with the results of our chemical 
evolution models adopting standard prescriptions for star formation, dust production and a standard, 
Salpeter IMF. 
The systems presenting detections for dust masses, SFRs, gas masses and dynamical masses 
represent the most luminous sources;  
our standard models fail to reproduce both the dust masses and the SFRs for these peculiar systems. 
\item To reproduce the high star formation rates at the observed gas masses 
of such peculiar systems, the star formation efficiency adopted in our standard models needs 
to be increased. This can be explained qualitatively by means of a ``positive'' AGN 
feedback, boosting star formation in the innermost regions of proto-spheroidals (Silk 2005). 
The enhanced SF efficiencies found in chemical evolution studies including this process support 
this conclusion (Pipino et al. 2009a). 
By increasing the star formation efficiency, the dust content is reduced owing to 
more efficient grain destruction by supernovae. 
\item The dust content of the systems with detections require an 
average dust yield per stellar mass larger than solar. 
This can be achieved by increasing the dust growth efficiency and by assuming a top-heavy IMF.
\item By assuming a dust growth efficiency 
an order of magnitude larger than the one implied by studies of the local dust budget, 
the discrepancy between our models and the observed dataset is reduced. 
Theoretical studies of dust production in starbursts indicate that, in extreme star formation conditions,  
larger growth efficiencies can be explained by larger average particle densities in high-z molecular 
clouds, as well as higher metallicities. 
\item Beside an enhanced star formation and dust growth efficiency, to fully account for 
the observed dust masses, a top-heavy stellar initial mass function is required. 
In this work, we have considered a Larson (1998) IMF with a 
a characteristic mass of $1.2$ M$_\odot$. 
A top-heavy IMF in high-$z$ starbursts can be explained by means of 
an average Jeans mass larger to the local one, due mostly to 
less efficient cooling and to  a particularly intense UV radiation field. 
A top-heavy IMF would imply a ten-times solar interstellar metallicity. No such vale 
has ever been observed so far, in any environment. In the future, 
the use of ALMA for the study of diagnostics such as the 
far-infrared fine-structure emission lines [NII] at 205 $\mu m$ and [CII] at 158 $\mu m$, 
already used for the analysis of the metal content of a massive starburst at z$\sim$ 4.8 (Nagao et al. 2012), 
will be of crucial importance to constrain the metallicity and hence the stellar IMF of high-redshift star-busts. 
\item The descendants of some of 
our systems may be potentially visible at lower redshift, in the form of large, passive objects. 
By means of a spectro-photometric code (Jimenez et al. 2004), starting from our model 
star formation history, we computed synthetic spectra 
for the descendants of the high-$z$ QSO hosts. 
In principle, the sensitivity limits of deep future surveys such as LSST and Euclid  
will allow one to reveal such systems already at large redshifts ($z>4$). 
By normalizing their number density to the BH mass function of high-$z$ QSO hosts, 
we predicted the expected number counts of such objects. Their presence could be 
significant in the wide Euclid survey, where $\sim 3.8 \, 10^5$ QSO host descendants 
are expected between $z=0$ and $z=2$. 
\end{itemize}
Future studies of the properties of QSO hosts will be very important 
to assess the nature of the progenitors of present-day ellipticals, and 
will serve as valuable benchmark for galaxy formation theories. 
As forthcoming work, it will be important to use the data collected in this work 
to investigate the dust content, 
the SF properties, and the buildup of both the bulge and black hole mass 
by means of semi-analytical galaxy formation models.

\section*{Acknowledgments}
We are grateful to A. Cimatti, L. Ciotti, C. Nipoti and G. Zamorani for 
several interesting discussions, and to an anonymous referee for valuable suggestions that improved the paper. 
We also thank R. Schneider for useful comments on an early version of the paper. 
FC and FM acknowledge financial support from PRIN MIUR 2010-2011, 
project "The Chemical and Dynamical Evolution of the Milky Way and Local Group Galaxies", 
prot. 2010LY5N2T. RG and CV acknowledge support from ASI-INAF grant I/009/10/0.

%
%
%
%
%
%
%

\label{lastpage}

\end{document}